
\overfullrule=0pt

\baselineskip 20pt
\parskip 0pt
\def \bk {\break}
\def \sect {\section}
\def \chap {\chapter}
\def \squiggle {\tilde}
\def \Lscr {{\cal L}}
\def \Hscr {{\cal H}}

\def\dslash{\not{\hbox{\kern-2pt $\partial$}}}
\def\Dslash{\not{\hbox{\kern-4pt  D }}}
\def\Aslash{\not{\hbox{\kern-4pt  A }}}
\def\TDslash{\not{\hbox{\kern-4pt $\tilde D$}}}
\def\Tdslash{\not{\hbox{\kern-2pt $\tilde \partial $}}}
\def\Lslash{\not{\hbox{\kern-2pt L}}}
\def\Qslash{\vert{\hbox{\kern-5pt Q}}}
\def\Rslash{\vert{\hbox{\kern-5.5pt R}}}
%

\def\pl#1{{\it Phys. Lett.} {\bf #1B}}
\def\prl#1{{\it Phys. Rev. Lett.} {\bf #1}}

\def\prd#1{{\it Phys. Rev.} {\bf D#1}}
\def\PR#1{{\it Phys. Rev.} {\bf #1}}
\def\np#1{{\it Nucl. Phys.} {\bf B#1}}



%
\def\fillbox#1{\hbox to #1{\vbox to #1{\vfil}\hfil}}
\def\dotbox#1{\hbox to #1{\vbox to 10pt{\vfil}\hss $\cdots$ \hss}}
\def\ggenbox#1#2{\vbox to 10pt{\vss \hbox to #1{\hss #2  \hss} \vss}}
%

%

%

%

%

%


\def\lsquare{\left[}
\def\rsquare{\right]}

\def\bold#1{\setbox0=\hbox{$#1$}%
     \kern-.025em\copy0\kern-\wd0
     \kern.05em\copy0\kern-\wd0
     \kern-.025em\raise.0433em\box0 }

\catcode`\@=11 
\def\lsim{\mathrel{\mathpalette\@versim<}}
\def\gsim{\mathrel{\mathpalette\@versim>}}
\def\@versim#1#2{\vcenter{\offinterlineskip
        \ialign{$\m@th#1\hfil##\hfil$\crcr#2\crcr\sim\crcr } }}
\catcode`\@=12 

\def\refmark#1{$^{[#1]}$}
\def\naive{na\"{\i}ve}
\def\ket#1{\left| #1\right\rangle}
\def\VEV#1{\left\langle #1\right\rangle}
\def\ie{{\sl i.e.}}

\def\journal#1&#2(#3){\unskip, {\sl #1} {\bf#2}(#3)}
%
\def\lf{\leaders\hbox to 1em{\hss.\hss}\hfill}
\titlepage

\rightline {WIS-92/54/June-PH}

\rightline{TAUP- 1981-92}

\rightline{hep-th/9207017}
\vskip 1cm
\title{Bosonization and QCD in Two Dimensions}
\author {Y. Frishman }
\address{ Department of Physics, Weizmann Institute of Science \break
 76100 Rehovot Israel}
\author { J.~Sonnenschein }
\address{ School of Physics and Astronomy\break
Beverly and Raymond Sackler \break
Faculty    of Exact Sciences\break
Ramat Aviv Tel-Aviv, 69987, Israel}
\bigskip
\bigskip
\endpage
\abstract{
Quantum Chromodynamics (QCD) is believed to describe
the strong interactions.
In the asymptotic domain of large momenta, improved perturbation theory
describes phenomena by use of point like quarks and gluons.
But the spectrum and
wave functions are in the non-perrturbative domain,
for which not much can be
done analytically in four dimensions.
In order to develop analytical methods
physicists turned to simpler models,
 like $QCD_2$, the theory in one-space and one-time dimensions.

 This review is
devoted to the application of bosonization
techniques to two dimensional QCD. We
start with  a description of the ``abelian bosonization".
The methods  of the
abelian bosonization are  applied to  several examples like  the Thirring
model, the Schwinger model and QCD$_2$.
The failure of this scheme to handle flavored fermions is explained.
Witten's non-abelian bosonization rules
are summarized including  the generalization to  the case of fermions
with color and flavor degrees of freedom.
We discuss in details the bosonic version of the mass bilinear of
colored-flavored fermions in various schemes.
The color group is gauged and the full bosonized version of massive
multiflavor QCD is written down.
The strong coupling limit is taken in  the
``product scheme"  and then in the
$U(N_F\times N_C)$ scheme.
Once the multiflavor $QCD_2$ action in the interesting region of the low
energies is written down, we extract
the semiclassical
 low lying baryonic spectrum.
First classical soliton
solutions of the bosonic action are derived. Quantizing the flavor space
around those classical solutions produces the masses as well as the
flavor properties of the two dimensional baryons. In addition
low lying
multibaryonic solutions  are presented, as
well as wave functions and matrix
elements of interest, like $q\bar q$ content.}
\endpage
\noindent{\bf Contents}\hfill\break
\noindent\line{1.\ Introduction\lf 4}

\line{2.\ Abelian Bosonization \lf 8}
\line{ 2.1 \ The Equivalence of the Free Massless Theories\lf 8}
\line{ 2.2 \ The Thirring Model\lf 11}
\line{ 2.3 \ Abelian Bosonization of the mass term\lf 12}
\line{ 2.4 \ The Massive Schwinger Model\lf 14}
\line{ 2.5  \ Bosonization of  Flavored fermions\lf 16}
\line{ 2.6 \  Abelian Bosonization of flavored $QCD_2$ \lf 18}
\line{3.\ Non-Abelian bosonization of colored flavored fermions\lf 21}
\line{ 3.1 \ Introduction\lf 21}
\line{ 3.2 \ Witten's Non-abelian bosonization\lf 21}
\line{ 3.3 \ Non-Abelian bosonization of Dirac fermions with color
and flavor \lf 24}
\line { 3.4 \ The bosonization of a mass bilinear of
  Dirac fermions \lf 29}

\line{ 3.5 \ Bosonization of a mass bilinears in the product scheme
   \lf 30}
\line { 3.6 \ Bosonization using the $U(N_F\times U_C)$ WZW action\lf 31}
\line{4.\  Bosonized Massive
Multiflavor $QCD_2$ \lf 35}
\line{ 4.1\ The ``Hybrid" approach\lf 35}
 \line{ 4.2\ Gauging the WZW action\lf 36}
\line{ 4.3\  The strong coupling limit\lf 39}
 \line{ 4.4\ Multiflavor $QCD_2$ using the $U(N_F\times N_C)$ Scheme\lf 40}
\line{5.\ The Baryonic Spectrum of Multiflavor $QCD_2$ \lf 43}
\line { 5.1\ Classical Soliton solutions\lf 43}
\line { 5.2 \ Semi-classical quantization and the Baryons\lf 45}
\line{ 5.3 \ The Baryonic spectrum\lf 55}
\line{ 5.4 \ Flavor quark content of the baryons \lf 56}
\line{ 5.5\ Multibaryons\lf 59}
\line{6.\ Summary and Conclusions\lf 64}
\line{7.\ List of references \lf 68}
\vfil\eject
\REF\Gell{ M. Gell-Mann {\it Physics } {\bf 1} (1964) 63.}
\REF\DaAd{ R. Dashen and S. Adler, ``Current Algebras and Applications
to particle physics", W. A. Benjamin Inc. (1968).}
\REF\Sug{ H. Sugawara, \PR 170 (1968) 1659.}
\REF\Som{ C. Sommerfield, \PR {176} (1968) 2019.}
\REF\BaFrHa{K. Bardakci, Y. Frishman and M. Halpern, {\it Phys. Rev.}
{\bf 170} (1968) 1353.}
\REF\DaFr{R. Dashen, and Y. Frishman, \pl {46} (1973) 439.}
\REF{\FD }{
R. Dashen and Y. Frishman, \prd {11} (1975) 2781\ .}
\REF\FrLC{ Y. Frishman, \prl {25} (1970) 966; Abstract to Kiev (April
1970), $XV^{th}$  Int. Conf. on High Energy Physics.}
\REF\FrGel{H. Fritzsch and M. Gell-Mann,
Coral Gables Conference on Fundamental Interactions at High Energies
1971 (Gordon \& Breach), page 1.}
\REF\FrPR{ Y. Frishman, {\it Phys. Reports} {\bf 136} (1974) 1,
and reference therein.}
\REF{\DFZ}{G. F.
Dell'Antonio, Y. Frishman and D. Z. Zwanziger, \prd{6} (1972) 988.\ }
\REF\tH { G. 't Hooft, \np  {75} (1974) 461.}
\REF{\Co}{S. Coleman, \prd       {11} (1975) 2088.\ }
\REF{\Man}{S. Mandelstam,
\prd {11}            (1975) 3026.}
\REF{\Halp}{M. B. Halpern, {\it Phys. Rev.} {\bf D12} (1975) 1684.}

\REF\CFG{E. Cohen, Y. Frishman and D. Gepner, \pl {121} (1983) 180.}
\REF{\Wi}{E. Witten, Comm. Math. Phys. {\bf 92} (1984) 455.}
\REF{\DFS}{G.D. Date, Y. Frishman and J. Sonnenschein,
\np {283} (1987) 365.}
\REF{\FS}{Y. Frishman and J. Sonnenschein,
\np {294} (1987) 801.}
\REF\YFWZ{Y. Frishman and W.J. Zakrzewski, in $25^{th}$
 Int. Conf. on High
Energy Physics, Singapore 2-8 August 1990, page 936.}
\REF\rMarek{Y. Frishman and M. Karliner,
\np {344}  (1990) 393.}
\REF{\Skyrme}{T.H. Skyrme \journal Proc.
Roy. Soc. London &A260 (1961) 171.}
\REF{\SM}{E. Witten, \np {223} (1983) 422,
{\sl ibid}  433;
G. Adkins, C. Nappi and E. Witten,\np
 {228} (1983) 433.
For the 3 flavor extension of the model see
E. Guadagnini \np {236} (1984) 35;
P. O. Mazur, M. A. Nowak and M. Prasza\l owicz,
 \pl {147} (1984) 137.}
\REF\RuCa{V. A. Rubakov, {\it Pisma Zh. Eksp. Teor. Fiz. }
{\bf 33} (1981) 658.
\break\hfill
C. G. Callan, \prd {25} (1982) 2141; {\bf D26} (1982) 2058.}
\REF\FGY{Y. Frishman, D. Gepner and S. Yankielowicz,
\pl {141} (1984) 76.}

\REF\ElFrKa{J. Ellis, Y. Frishman and M. Karliner, CERN preprint
CERN-TH 6426/1992.}
\chapter {Introduction}
In this review  we discuss methods of writing theories with fermions in
terms of bosonic variables only, with the aim of arriving as close as
possible to analytic solutions of Quantum Chromodynamics (QCD). So far,
except for some very special situations involving very massive objects
(see below), these methods have not been applicable to four dimensional
space-time, but have been very successful in the case of two
dimensions.

Bosonic variables describe
 various interactions that involve
bosons as well as fermions. The interactions of the standard model are
all carried by currents which are bosonic variables. The electro-weak
and strong interactions are carried by $SU_W(2)\times U(1)$ and
$SU_C(3)$ currents.
In fact, even before QCD, the importance of weak and electromagnetic
currents was realized in terms of Gell-Mann's current
algebra.\refmark{\Gell,\DaAd} This led the way to attempts at expressing
the strong energy-momentum tensor in terms of these
currents\refmark{\Sug,\Som}, and it turned out to be   as a
quadratic form. These approaches were extended to include
electromagnetic and other interactions.\refmark\BaFrHa\
 However, the coefficient\refmark\Sug\  in the energy-momentum tensor was
proportional to  the inverse
of $k$, the Kac-Moody level. In four dimensions
the analog is infinite and cut-off dependent.
The handling  in four dimensions
was not precise. The situation was clarified when two-dimensional models
with a non-abelian symmetry group were considered.
It was then found that the
coefficient
actually  is proportional to the inverse of  $k+C_G$,
where $C_G$ is the second
Casimir operator of the adjoint representation,
 rather then to the inverse of
 $k$.\refmark{\DaFr,\FD}
This came out after a careful handling of the operator products by
normal ordering. The exact expression for the energy momentum tensor in
terms of currents,
first found in ref. [\DaFr] for $SU(N)$ (eqs.(5)\&(12) there), is being
widely used in conformal field theory as well as string theory.

 Sometime before
that, operator product expansions near the light-cone were investigated
\refmark{\FrLC,\FrGel,\FrPR}. Products of currents, in this domain,
were instrumental in relation with deep-inelastic scattering of the
leptons $(e,\mu, \nu)$ off hadrons.
Afterwards, the
Thirring model was solved in terms of currents\refmark\DFZ\
signaling the possibility of having fermionic theories in terms of
bosonic variables.

An important step forward in solving non-abelian gauge theories was
achieved by 't Hooft,\refmark\tH
 who found the meson spectrum in $QCD_2$ in the
limit of large number of colors $N_C$, with $e^2_c N_C$ fixed. (Many
works on that followed, but this is not part of our review). The baryon
spectrum, however, remained unsolved untill the methods of bosonization
were developed.

The start of bosonization was with the now called ``abelian
bosonization" of
Coleman\refmark\Co and Mandelstam,\refmark\Man generalized to
many fermions by Halpern.\refmark\Halp\break
Attempts of solving $QCD_2$ using these
schemes met with technical difficulties.\refmark\CFG\
A fundamental step forward was obtained by Witten\refmark\Wi  introducing
``non-abelian bosonization". This opened the way to analyzing $QCD_2$ in
the semi-classical limit obtaining the
low-lying baryon spectrum,\refmark{\DFS,\FS}
multibaryons\refmark\YFWZ
 and matrix elements of interest,  like for example various
quarks content.\refmark\rMarek
The treatment is in the spirit of Skyrme model,\refmark\Skyrme
 which in four dimensions
is an ``educated guess" of the low energy effective action.\refmark\SM
A full
discussion and references are in the review.

Bosonization techniques have been widely
applied and developed in the context
of string and superstring theories.
The applications varied from bosonization
of world-sheet fermions in various superstring
theories through the bosonic
construction of fermion vertex operators and
all the way to the bosonization of
anticommuting and commuting ghost
systems of dimension $\lambda,1-\lambda$. The
 bosonic description in the heterotic string motivated the discussion of
bosonization of chiral fermions. Various formulations were suggested for
chiral bosons and a large body of
literature deals with this topic. Another
 development important for string
 theories was the derivation of bosonization
rules for higher genus Reimann surfaces.
The action of the
``non-abelian bosonization" which is now referred to as the
WZW model is a very essential  tool in the realization of
rational conformal fields theories which relate to the Kac-Moody algebras.
Gauged WZW models
were invoked to describe conformal coset space  models and
led to interesting string models like the two dimensional black hole
background.
Since all these string and conformal field theory related topics
are  a large subject
 matter by themselves
 and since they are not directly related to the topic of
the present review, they  are not discussed  here and the reference list
does not cover them.

As for the four dimensional cases, the treatment was in connection with
monopole induced proton decay\refmark\RuCa
 and fractional charges induced on
monopoles by light fermions.\refmark\FGY\
 In these cases the light fermions move in
the field of a very heavy particle, so that only the lowest wave is
considered, thus obtaining dynamics in the radial variables only,
namely, one space dimension. One can then use the bosonization
techniques. To get a theory on a whole $(-\infty, \infty)$ line, a
``reflection principle " was used\refmark\FGY
 as the boundary conditions at $r=0$ were
suitable.

We hope that some of methods discussed in this review will find
application in four dimensional physics, as some feature of the quark
content and spectrum indicate.
Let us also mention recent work\refmark\ElFrKa ( that we will not
review), to get the `` constituent quarks" out of the basic $QCD_2$
Lagrangian.

This  review is organized as follows. Chapter 2 is devoted to  abelian
bosonization. We start with  Mandelstam's bosonic expressions for
 free
massless Weyl fermions. The associated chiral currents and energy
momentum tensor are discussed via the corresponding Kac-Moody and
Virasoro algebras. The ideas of bosonization are  applied to  the
examples of the Thirring  model and then to
massive fermion. We also describe
the coupling of the bosonized theory to an electromagnetic field, namely,
the Schwinger model. Next we introduce
 flavor degrees of freedom and mention some
 related problems. Finally the abelian
bosonization of flavored QCD is written down and discussed.
The second topic,  presented in chapter 3, is the non-abelian bosonization
of colored-flavored fermions. Witten's non-abelian bosonization rules
are summarized followed by the generalization for the case of fermions
with color and flavor degrees of freedom.
Whereas a consistent bosonization  of  the mass term of unflavored
fermions is quite straightforward, the analog for both color and flavor
groups, faces some difficulties. It is shown that the bosonic mass term
in what is  referred to as the ``product scheme" fails to reproduce
fermionic correlation functions. We then discuss a solution to this
difficulty via the bosonization of
$U(N_F\times N_C)$ WZW action at level $k=1$.
Chapter 4 deals with the bosonic version of massive multiflavor $QCD_2$.
We first describe a ``Hybrid" approach where abelian bosonization is
invoked to take care of the color degrees of freedom and a WZW picture
to cope with the flavor ones. Coming back to the full non-abelian
bosonization approach, two prescriptions of gauging  the corresponding
$WZW$ action are described. The strong coupling limit is taken in  the
``product scheme"  and then in the
$U(N_F\times N_C)$ scheme.
Once the multiflavor $QCD_2$ action in the interesting region of the low
energies is written down, we devote chapter 5 to
the semiclassical
extraction of the low lying baryonic spectrum. Classical soliton
solutions of the bosonic action are derived. Quantizing the flavor space
around those classical solutions produces the masses as well as the
flavor properties of the two dimensional baryons. In addition
low lying
multibaryonic solutions  are presented.
We also review baryonic wave functions and matrix elements of interest,
like various quarks content.
Chapter 6 includes a summary of the review together with some concluding
remarks and some open problems.

\centerline {Notations}

We use Minkowski metric with $g_{00}=1$ and $g_{11}=-1$.
Light-cone coordinates are written as
$x_\pm = {1\over \sqrt{2}}(x_0\pm x_1)$, and
$\epsilon_{01}=1$.

$\half T^A$ are the matrices representing $SU(N)$ generators, with
$Tr(T^AT^B)=2\delta^{AB}$.

$e_c$ is the $QCD_2$ gauge coupling. It has a dimension of mass, thus
being the analog of the $QCD$ scale in four dimensions.

The constant $c$ appearing in the bosonization formulae is $c=\half
e^\gamma\simeq
0.891$, where $\gamma$ is the Euler constant. $\tilde c$ is
given by $\tilde c={c\over 2\pi}$.
\ack{
We are greatful to Dr. Eyal Cohen for his help at the very early
stages  of the review. One of us (J.S.) would like to thank ``the Einstein
Center for Theoretical Physics"
of the Weizmann Institute for its support.
\endpage
\def\({\lbrack}
\def\){\rbrack}
\def\p{\partial}

\chapter {Abelian  Bosonization}
\medskip
The heart of the bosonization idea is the equivalence of the massless
Dirac field and the massless scalar field in two space-time dimensions.
There is a correspondence at the operator level between all operators of
one theory and those of the other.  This correspondence allows one to
relate also the mass term and  possible interaction terms and thereby
to extend the equivalence beyond the free massless level.
\section {The Equivalence of the Free Massless Theories}
\medskip
\par
The explicit construction of
fermion fields in terms of  boson
fields is due to Mandelstam.\refmark\Man\
 Left and right Weyl fermions
 $\Psi_L$and $\Psi_R$
 are given by
$$\eqalign{\Psi_L&=
\sqrt{{c\mu\over 2\pi}} : {\rm exp}
\bigg\(-i\sqrt \pi\bigg (\int\limits^x_{-\infty}d\xi\pi(\xi)+\phi
(x)\bigg )\bigg\) :\cr \Psi_R&=
\sqrt{{c\mu\over 2\pi}} : {\rm exp}\bigg
\(-i\sqrt\pi\bigg (\int\limits^x_{-\infty} d\xi\pi(\xi)-\phi(x)\bigg
)\bigg\) : \cr}\eqn\mishza$$
where $c$ is a constant.  A computation yields $c=\half e^\gamma\sim
0.891$, where $\gamma$ is the Euler constant. The
normal ordering denoted by :  : is
performed with respect to the scale $\mu$.
\par
The equal time commutation relations of the $\phi$-field
$$\bigg\(\phi(x,t),\pi (y,t)\bigg \) = i\delta (x-y) \eqn\mishab $$
imply the canonical anti-commutation relations for the $\Psi$ field:
$$\{\Psi ^\dagger_{L,R} (x,t),\Psi_{L,R}(y,t)\} = \delta (x-y) \ .
\eqn\mishaab$$
as can be verified using the explicit construction eq. \mishza .
The fermion
field $\Psi$
is therefore, an inherently non-local functional of the scalar
field. However fermion bilinears, such as currents or masses, are local
functionals.  The vector current, for example
$$J^\mu = :\bar\Psi\gamma^\mu\Psi : = {-1\over\sqrt\pi} \epsilon^{\mu\nu}
\partial_\nu\phi\eqn\mishaa$$
This
identificination of $J^\mu$ leads automatically to a conserved current
$$\partial_\mu J^\mu = 0$$
independent of the equations for $\phi$. This is a ``topological"
conservation, connected with choosing the ``vector conservation" scheme
(see also    3.2).
In the later applications, we will demand more freedom in the
renormalization of interacting theories, in particular the possibility to
have a vector current anomaly.  The bosonization procedure there will
therefore be somewhat modified.  The modification will correspond to a
change of renormalization scheme.
\par
The overall coefficient of the current is  such that the
fermion number charge
$$Q = \int\limits^\infty_{-\infty} dx j_0 (x)= 1 \eqn\mishae $$
 for the $\Psi$-field.  This is the normalization condition that we
will follow throughout.

The equivalence of the bosonic and fermionic descriptions is manifested
in the fact that the two theories have the same current algebra and the
same Virasoro algebra for the energy-momentum.
 The later is constructed from
the currents in a Sugawara form.

Let us start with the current algebra.
In addition to the ``topologically" conserved vector current the bosonic
theory has an axial current $J_\mu^5={1\over\sqrt{ \pi}}\partial_\mu
\phi$ which is the Neother current associated with the invariance of the
bosonic action under the global shift $\delta\phi=\epsilon$. One can
then define the left and right chiral currents $J^\mu_\pm = J^\mu \pm
J^\mu_5$ which correspond to shifts with $\epsilon(x_+)$ and
$\epsilon(x_-)$. Using the commutation relation
eqn.\mishab\  the following current algebra is
found
$$\bigg\(J_\pm (x_\pm), J_\pm (x'_\pm)\bigg\) = {2i\over \pi}
\delta' (x_\pm-x'_\pm)\eqn\mishzac$$
This is the same algebra as that of the fermionic chiral currents.
The algebra of the currents and the energy-momentum tensor  take a more
familiar form  once expressed in terms of the Laurant modes of
operators. Passing to the euclidean plane, using complex coordinates
$z=x+it$, and defining
$J=\sqrt{\pi} J_- , \bar J=\sqrt{\pi} J_+$
one imediately realizes that the  later currents are holomorphic and
anti-holomorphic functions respectively $\bar\p J =\p \bar J =0$.
Expanding $J$ in a Laurant series  $J=\sum_n J_n z^{-(n+1)}$ leads to
the Kac-Moody form of the  current algebra
$$ [J_n,J_m] =+{i\over 2}n\delta_{n+m,0} \eqno{\eq}$$
and a similar algebra for $\bar J$.
The bosonic energy-momentum tensor which is constructed from
the chiral currents $T_{\pm\pm} = \pi  :J_\pm J_\pm :$ obeys the
identical Virasoro algebra
$$\bigg\(T_\pm(x_\pm), T_\pm(x'_\pm)\bigg\) =
2i\bigg\(T_\pm(x_\pm)+T_\pm(x'_\pm)\bigg\)\delta'(x_\pm-x'_\pm)
-{i\over 6\pi}\delta'''(x_\pm-x'_\pm)\eqn\mishaar$$
which is identical to that of the fermionic energy-momentum tensor.
Translation to the Laurant modes
  $T=\sum_n L_n z^{-(n+2)}$ leads to
leads to the well known form of the Virasoro algebra
$$ [L_n,L_m] =(m-n)L_{m+n}+
{1\over 6}(m-1)m(m+1)\delta_{n+m,0} \eqno{\eq}$$
Coupling the theory to an abelian gauge field, the equivalence of the
bosonic and the fermionic formulations can be demonstrated by showing
that the corresponding effective actions are identical. This is
presented in section 3 for the non-abelian case.
\par
The equivalence of the bosons and the fermion bilinears is not only
mathematical.  The fermion Fock-space contains those bosons as physical
states.\refmark{\DFZ,\Co}
The reason for this is that in one
space dimension a massless field can move either to the left or to the
right.  A Dirac fermion and its anti-particle having together zero
fermionic charge and moving in the same direction will never separate.
They are therefore indistinguishable from a free massless boson.  This
picture changes when masses are introduced, and
the above relations will be approached at energies high compared
to the mass scale.
\section {The Thirring Model}
\par
The Thirring model is a current-current interaction, given by the
Lagrangian density\REF{\Thi}{W. Thirring, {\it Ann. Phys.} (N.Y)
{\bf 3} (1958) 91}\refend
$$\Lscr = i\bar\Psi \dslash\Psi - {1\over 2} g J^\mu J_\mu \eqn\mishaba$$
where $J_\mu ={:\!\bar\Psi\gamma_\mu\Psi\!:}$\ .
The model is exactly solvable \REFS{\Johnson}{K. Johnson, {\it Nuovo
Cimento} {\bf 20} (1961) 773.}\REFSCON{\Klai}{B. Klaiber, in Lectures
in Theoretical Physics XA (1968) 141 (Gordon and Breach, New York).}
\refsend
 and meaningful for $g>-\pi$.  Dell'Antonio, Frishman and
 Zwanziger\refmark\DFZ\
studied the model further by means of the operator product
expansion on the light-cone.  They expressed the fermionic bilinears of
the model as function of the current, and obtained expressions which seem
very natural in the light of the bosonization procedure which was
discovered later.
The limit of $g>-\pi$ ( see also
eq. (2.13)  )    follows from the scheme $a=1$ in
ref.\(\DFZ\), namely that $\Psi$ has a vector charge one.
\par
Following Mandelstam\refmark\Man
, we generalize the bosonization formula (1)
to include a parameter $\beta$, whose role will become clear later.
$$\eqalign{
\Psi_L &= \sqrt{c\mu\over2\pi} : {\rm exp}\bigg \(
-i\sqrt{\pi}({2\sqrt{\pi}\over
 \beta }\int\limits^x_{-\infty} d\xi\pi (\xi) +
 {\beta\over 2\sqrt{\pi}} \phi
(x))\bigg \):\cr
\Psi_R &= \sqrt{c\mu\over2\pi} : {\rm exp}\bigg \(
-i\sqrt{\pi}({2\sqrt{\pi}\over
 \beta }\int\limits^x_{-\infty} d\xi\pi (\xi) -
 {\beta\over 2\sqrt{\pi}} \phi
(x))\bigg \):\cr}
\eqn\mishag $$
\par
The equal-time
anti-commutation relations are still obeyed.  Eq. (3) for the vector
current is now modified to:
$$J^\mu =- {\beta\over 2\pi} \epsilon^{\mu\nu}\partial_\nu\phi \ , \
\eqn\mishagg$$
where the normalization corresponds again to the charge being the
fermion number appropriate to the field $\Psi$.  Eq. \mishag\
provides the operatoric
solution to the Thirring model where $\Psi$ is a free massless field.
The constant $\beta$ is related to the coupling constant $g$ through the
formula\refmark\Co
$${\beta^2\over 4\pi} = {1\over 1+g/\pi}\eqn\mishaggg$$
Eq. \mishagg\ implies that the central term in the Kac-Moddy algebra will
be ${\beta^2 \over 4\pi^2}$,
  as compared to the ${1\over \pi}$ of the free case.
\par
The analogy between the fermionic and  bosonic theory therefore holds
even when an interaction is introduced.  The equivalent bosonic theory is
still free, however.
\section{ Abelian Bosonization of the mass term}
We now introduce a mass term for the fermion.  The mapping to a bosonic
theory still holds.  Moreover, its form is precisely (2.15) below.
  The
bosonic field will now be interacting, and the interaction will be
precisely the mapping of the fermionic mass term.
\par
The definition of the mass term, as that of the current, requires some
care due to the appearance of the products of operators at the same
point.  In fact, when $x$ approaches $y$ one gets\refmark\Man\
the operator product expansion
$$\eqalign{\Psi_R^\dagger (x)
 \Psi_L(y) &={c\mu\over 2\pi}|c\mu(x-y)|^\delta      :
e^{-i\beta\phi}:\cr
\Psi_L^\dagger(x) \Psi_R(y)&=
{c\mu\over2\pi}|c\mu(x-y)|^\delta : e^{i\beta\phi}:\cr}
\eqn\mishah $$
where $\delta = - {g\over 2\pi}(1+{\beta^2\over 4\pi})$.
The proper fermion mass term will therefore be defined by
$$ lim_{y\rightarrow x}
\int\limits^\infty_{-\infty} dx\
|c\mu(x-y)|^{-\delta}m\bar\Psi(x)\Psi(y) =
{c\mu\over\pi}m\int\limits^\infty_{-\infty} dx : {\rm cos}\beta\phi (x):
\eqn\mishai $$
With $\mu$ chosen such that
$m={\mu\pi\over c\beta^2}$, the mass term transforms
in the bosonic language to
$$\Delta \Lscr = {\mu^2\over\beta^2}: {\rm
cos}\beta\phi:$$ We remind the reader that the normal ordering is
performed with respect to $\mu$.
\par
The bosonic model at which we arrived is the Sine-Gordon
interaction.  The classical Sine-Gordon model admits a soliton solution.
It is time-independent and interpolates between adjacent wells of the
scalar potential with finite energy.  In the quantum theory this
classical solution becomes a particle. (For a review of the Sine-Gordon
system
\REF\Raj{R.  Rajaraman, {\it Phys.  Rep.} {\bf 21} (1975) 227\ .}
see ref.\(\Raj\). The
classical  equation of motion is
$$\phi" -   {\mu^2\over\beta} {\rm sin} (\beta\phi) =
0\eqn\misham$$
The  static soliton solution is given by
$$\phi = {4\over\beta} {\rm tan}^{-1}[{\rm exp} \mu(x-x_0)]$$
where $x_0$ is the ``center" of the soliton.
\par
Using eq. \mishae\
    we can compute the fermionic number of this solution.  It is
given by $$Q = {\beta\over 2\pi}\big\(\phi (\infty)-\phi
(-\infty)\big\) = 1\eqn\mishak $$ The fermion has therefore a direct
physical interpretation as the Sine-Gordon soliton.
\par
It is interesting to note a remarkable property of eq. \mishaggg\
 which relates
the coupling constants of the Thirring model and its bosonic equivalent,
the Sine-Gordon model. The weak coupling of one theory is the strong
coupling of the other.  This property occurs often in bosonized theories
and hints at the usefulness of the method in dealing with theories at
strong coupling, where perturbative methods tail.
\REF\CJS{S. Coleman, R. Jackiw and L. Susskind, {\it Ann. Phys.} {\bf 93}
  (1975) 267\ .}\REF{\COA}{
S. Coleman, {\it Ann. Phys.} {\bf 101} (1976) 239\ .}
\section {The Massive Schwinger Model}
\par
The bosonization technique has turned out to be a useful tool in the
investigation of other two dimensional models.  Our next example is the
massive Schwinger model.\refmark{\CJS,\COA}
 The Lagrangian of this model is given by
$$\Lscr = - {1\over 4} F_{\mu\nu} F^{\mu\nu} +\bar\Psi
(i\dslash - e\Aslash- m)\Psi\eqn\mishaka $$
As is well-known, the gauge field in two space-time dimensions is not
dynamical.  Following Coleman\refmark\COA\ we choose the axial gauge
$A_1=0$, and
the other component $A_0$ can be solved as a function of the electric
current.  The resulting electric field will be
$$F_{01} = e\partial_1^{-1} j_0 + {e\theta\over 2\pi}\qquad J^\mu = :
\bar\Psi\gamma^\mu \Psi :\eqn\mishakb$$
$\theta$ is a new parameter in the theory.  It is a vacuum angle,
analogous to the vacuum angle due to instanton
tunnelling in 4-dimensional
QCD.
It is the conjugate to the winding number, appearing in two dimensions
for the abelian case,
 as follows from $\Pi_1\(U(1)\) = {\cal Z}$ ( looking at a
circle of large radius in the two dimensional plane).
Physics is invariant under $\theta\rightarrow\theta +2\pi$. From the
expression above    it is clear that ${e\theta\over 2\pi}$
corresponds to a background electric field.  The periodicity is due to
the ability to produce electron-positron pairs is the vacuum when
$|{e\theta\over 2\pi}|>{1\over 2} e$, and these pairs create their own
electric field which reduces the original one.
\par
The Hamiltonian density in our gauge is
$$\Hscr =\bar\Psi (i\gamma_1\partial_1+m) \Psi + {1\over 2}
(F_{01})^2\eqn\mishakc $$
Bosonizing the current according to eq. \mishaa\  we find
$$\Hscr =
:\bigg\({1\over 2}\pi^2 +{1\over 2}(\partial_1\phi)^2 -{cm^2 \over
\pi}cos(2{\sqrt\pi}\phi)
+{e^2\over2\pi}\bigg ({1\over 2} {\theta\over\sqrt\pi} -
\phi\bigg )^2\bigg\):\eqn\mishakd$$
where the normal ordering is with respect to the mass $m$.
\par
After a shift in the definition of $\phi$ $$\phi\rightarrow\phi +
{1\over 2}{\theta\over\sqrt\pi}\eqn\mishakf$$
and normal ordering with respect to
$\mu^2=e^2/\pi$ one finds
$$\Hscr = :\bigg\( {1\over 2}\pi^2 +{1\over 2}
(\partial_1\phi)^2 +{1\over 2}\mu^2\phi^2 -
{cm\mu \over \pi}
{\rm cos} (\theta + 2\sqrt\pi\phi)\bigg\):\eqn\mishakg$$
{}From this expression the periodicity in
$\theta$ is manifest.  In the strong coupling limit, the bosonized form
of the Hamiltonian is very useful.  The theory contains a meson of a mass
that is approximately $\mu$, and depending on the value of $\theta$,
the number of excitations of that meson is zero, finite or infinite.

For convenience, we will denote
$$\tilde c={c\over \pi}\eqn\mishbtc$$
from here on.
\section{Bosonization of  Flavored fermions}
\par
So far we have treated the case of one fermion flavor.  The several
flavor case is in fact more subtle than it appears at first sight, due to
the different realization of global symmetries in the fermionic or
bosonic languages.
In the next chapter a ``natural" scheme for bosonizing flavoured
fermions, the ``non-abelian bosonization",
will be introduced. Here we review the ``abelian bosonization" attemps.
\par
Flavor was first introduced in the bosonization
procedure by Halpern.\refmark\Halp
     The straightforward application of the procedure in eq. (2.1)
 for each one of the flavors will result in commuting
 fields $\Psi$ for
different flavors.  They can be made anti-commuting by the introduction
of a phase factor $K$ defined as a function of the fermion fields on
which it acts.
$$K_i = \prod_{j<i} (-1)^{n_j}\eqn\mishal $$
where $i,j = 1, \cdots, N_F$ are flavor indices. $n_j$ is the number of
fermi fields with index $j$ on which $K_i$ acts.  The bosonization
formula eq. (1) is modified to
$$\eqalign{\Psi^i_L &=\sqrt{c\mu\over 2\pi}K_i :
{\rm exp}\bigg\(-i\sqrt\pi\bigg
(\int\limits^x_{-\infty} d\xi\dot
\phi^i(\xi)+\phi^i(x)\bigg )\bigg\):\cr
\Psi^i_R&=\sqrt{c\mu\over2\pi}K_i : {\rm exp}\bigg\(-i\sqrt{\pi}\bigg
(\int\limits^x_{\infty} d\xi\dot
\phi^i(\xi)-\phi^i(x)\bigg )\bigg\):\cr}
\eqn\misha $$
This method of bosonization, when applied to the flavored Thirring model,
for example, transforms a fermionic system whose symmetry is
$U(N_F)$ to a
bosonic system whose manifest symmetry is only $\(U(1)\)^{N_F}$,
corresponding to adding constant values to the $\phi$ fields and
discrete $P_{N_F}$.  The scalar fields transform non-linearly under the
isopin, as induced from
eqs. \misha\
and the linear transformation
properties of the $\Psi$s. The isopin
quantum numbers of the soliton solutions are of
topological nature.\refmark\Halp\
In the next chapter we will discuss the non-abelian $U(N_F)$
bosonization, where the full group structure remains manifest.
\par The
flavored Thirring model, both massless and massive, was thoroughly
investigated by Dashen and Frishman.\FD
  In particular, they derived a
``Sugawara-like" expression for the energy-momentum tensor in terms of
the currents.  They separated the energy-momentum tensor into a singlet
part $T_{\mu\nu}^{(B)}$, and an $SU(N_F)$ part $T_{\mu\nu}^{(V)}$.
$$\eqalign{T_{\mu\nu}&=T_{\mu\nu}^{(B)} + T^{(V)}_{\mu\nu}\cr
T_{\mu\nu}^{(B)}&= {1\over 2\tilde C_0}\bigg\(2 : J_\mu J_\nu : -g_{\mu\nu}
: J_\lambda J^\lambda :\bigg\)\cr
T_{\mu\nu}^{(V)}&= {1\over 2\tilde C_1}\bigg\(2 : J_\mu^A J_\nu^A
: - g_{\mu\nu}:J_\lambda^AJ^{A\lambda}:\bigg\)\cr}\eqn\mishaqa$$
$\tilde C_0, \tilde C_1$ are constants.
\par
$\tilde C_0=C_0$ is the coefficient of the Schwinger term in the
equal-time continuation relation of the single vector current.  In light
cone coordinates
$$\bigg\(J_\pm (u), J_\pm (u')\bigg\) = 2iC_0\delta' (u-u')
\eqn\mishaqb$$
It is equal to
${N_F\over\pi}$ in the free field case, and in general serves as
a normalization of the current.
$\tilde C_1 = {N_F\over 2\pi}+C_1$, where
$C_1$ is the coefficient of the Schwinger term in the equal-time
communication relation of the $SU(N_F)$ vector currents.
$$\eqalign{\bigg\(J^A_\pm(u), J^B_\pm(u')\bigg\) &= 2i f^{ABC}
J^C_+(u)\delta (u-u') + \cr
&+2i C_1 \delta^{AB}\delta' (u-u') \cr}\eqn\mishaqc$$
$f^{ABC}$ being the $SU(N_F)$ structure constants.
It turns out that $C_1={1\over 2\pi}$ and thus
$$\tilde C_1 = {N_F+1\over 2\pi}\eqn\mishaqqc$$
The singlet energy momentum tensor equal-time commulates contains
Schwinger-type anominaly due to the normal ordering
$$\bigg\(T_+^{(B)}(u), T^{(B)}_+(u')\bigg\) =
2i\bigg\(T^{(B)}_+(u)+T^{(B)}_+(u')\bigg\)\delta'(u-u')
-{i\over 6\pi}\delta'''(u-u')\eqn\mishar$$
So does the $T^{V}$, with a factor of  $(N_F-1)$ multiplying the
$\delta'''$ factor.
The coefficient ${1\over 2\tilde C_1}$ in eqn.\mishaqa\
for $ T^{(V)}_{\mu\nu}$,
was first obtained correctly in ref.\(\FD\). The original expression of
Sugawara\refmark{\Sug,\Som}
did not include the part $N_F$ in $\tilde C_1$,
which comes froom carefull normal ordering of the products of currents
in the expressions for the energy-momentum ( what comes up is actually
the quadratic Casimir of the adjoint representation, which for $SU(N_F)$
is equal to $N_F$). Eqs.  \mishaqa\ are of basic use in conformal field
theories as well as string theories.
\section{ Abelian Bosonization of flavored $QCD_2$}
\par Let us now apply the prescription fo flavored Dirac fermions
for the  analysis of $QCD_2$. It is convenient to start with the
following Hamiltonian
\REF{\Bal}{V. Baluni, \pl  { 90} (1980) 407\ . }\refend\REF\St{
P.J. Steinhart, \np  {176} (1980) 100.}\refend
$$ H=(e_c)^2 \sum_{a,b=1}^{N_C} (E^a_b)^2 +
\sum_{a,b=1 }^{N_C}\sum_{i=1}^{N_F}
\bar\Psi^{ai}\gamma_1(i\delta_a^b\partial_1
-A^b_a)
\Psi_{bi} +m
\sum_{a=1 }^{N_C}\sum_{i=1}^{N_F}
\bar\Psi^{ai}\Psi_{ai}\eqn\mishas$$
in the gauge
$$ A_0=0 ; \ \ \ A^a_b = 0\ for\   a=b; \ \  E^a_b =0\ \ for\ a\ne b
\eqn\mishat $$
The Gauss law of the system is given by
 $$\partial_1 E^a_b =i\(A , E\)^a_b +\half\sum_{i=1}^{N_F}
 \Psi^{\dagger ai}\Psi_{bi} - {\delta^a_b
 \over 2N_C}
 \sum_{i=1}^{N_F}\sum_{d=1}^{N_C}
 \Psi^{\dagger di}\Psi_{ di}
 \eqn\mishatt$$
Bosonizing now the various parts of the Hamiltonian one then gets :
$$\eqalign{ H&=H_\Psi^0 + H_E -H^I\cr
H_\Psi^0 &= \Sigma_{ai}\left\( \half\(\pi^2_{ai} +(\partial_1
\phi_{ai})^2\)   +{cm\mu\over
\pi}:(1-cos2\sqrt{\pi}\phi_{ai}):\right\)\cr
H_E &={e_c^2\over 8\pi
N_c}\sum_{ab}\(\sum_i(\phi_{ai}-\phi_{bi})\)^2\cr
H^I &={2c^2\mu^2\over \pi^{3\over2}}\Sigma_{a\ne b}\Sigma_{ij}
K_{ij,ab}N_\mu
\(cos\sqrt{\pi}\int_{-\infty}^x(\pi_{ai}-\pi_{aj}+\pi_{bj}-\pi_{bi})(\xi)d\xi\)
\cr
&\(sin\sqrt{\pi}(\phi_{ai}+\phi_{aj}-\phi_{bj}-\phi_{bi})(\xi)\)
 \( \sum (\phi_{ka}-\phi_{kb})\)^{-1}\cr}\eqn\mishau$$
$H_\Psi^0$ is the free ``fermionic" part, after bosonization,
thus in terms of bosonic variables.
$H_E$ is the first term of the Hamiltonian eqn. \mishas\
rewritten in terms of the boson variables  corresponding to the
fermions, by eliminating the electric fields through the Gauss law. Thus
although originally coming from the kinetic part of the gauge
potentials, it actually involves  the interactions.
This is a result of the fact that there are no transverse vectors in
$1+1$ dimensions.
$K^{ab}_{ij}$ is the
generalized ordering operator (below).
\par In the case of one flavor,
$i=j=1,\ H^I$ does not involve  the  $\pi$ variables. This is the case
that first appeard in [\Bal].
           The generalization to many flavors, as
in ref.\(\St \), did not have the $H^I$ correctly ( for example the
$\pi's$  did not appear at all). The full expression first appears in ref.
\(\CFG\) , where the reader can also find  the definition of
$K^{ab}_{ij}$ ( generalizing the $K_i$ of eqn.\misha\ ).
\par The interaction term involves non-local terms which relate to colour
non-singlets.
For static and $e_c\rightarrow \infty$
approximation one            finds\refmark\CFG\
 that for $N_F=2$ the
interaction is field independent. For       $N_F\ge 3$, on the other
hand, the limit is singular. This singularity should    not be there in
the predictions of physical quantities, but it renders further treatment
very complicated.\refmark\CFG
\par It is thus clear that a different method of bosonization is required
for   the treatment of flavored $QCD_2$.
In the following chapters it will be shown that  the so called
``non-abelian   bosonization" which was introduced by
Witten\refmark\Wi
 is an adequate tool  for this purpose.
\endpage

\def \chap {\chapter}

\def \bk {\break}
\def \Lscr {{\cal L}}
\baselineskip 24pt
\parskip 0pt
\chap {Non-Abelian Bosonization of colored flavored  Fermions}
\section {Introduction}

\par
\par The non-abelian bosonization  introduced by Witten
is a set of rules
assigning    bosonic operators to fermionic ones, in a
theory of free
fermions invariant under a  global
non-abelian symmetry.
In \(\Wi \) the symmetry considered was $O(N)$
The bosonic operators are not expresed in terms of free bosonic fields like in
the abelian bosonization but rather in terms of interacting group elements.  In
particular, bosonic expressions can be written for the energy-momentum tensor,
the various chiral currents, the mass term and the complete action.

\section {Witten's Non-abelian bosonization}

 \par Let us start with
  $N$ free Majorana fermions  governed by the
action $$ S_{\Psi } = {i\over 2}\int d^2x
(\Psi_{-k}\partial_+ \Psi_{-k} +
\Psi_{+k}\partial_- \Psi_{+k} ) \eqno{\eq} $$
where $\Psi_- ,\Psi_+$ are left and right Weyl-Majorana spinor fields,
$\partial_{\pm}={1\over \sqrt{2}}(\partial_0 \pm \partial_1)$ and
$k=1 ,....N$.
The corresponding bosonic action is the  Wess-Zumino-Witten
(WZW) action:
$$ \eqalign{\squiggle S[u]=&{1\over 2}S[u]\cr
S[u]=&{1\over 8\pi}\int d^2xTr(\partial_\mu u\partial^\mu u^{-1})
+ {1\over12\pi}\int_B d^3y\varepsilon^{ijk}Tr(u^{-1}\partial_iu)
(u^{-1}\partial_ju)(u^{-1}\partial_ku)\cr}  \eqn{\mishcb}$$
where $u$ is a matrix in $O(N)$ whose elements are
bosonic fields.
The second term, the
Wess-Zumino
(WZ) term,\REF{\Fg}{J.  Wess and B.  Zumino, \pl {37} (1971) 95.}
\refend\  is defined on the ball B whose boundary S is taken to be
the euclidian two dimensional
space-time. Now, since $\pi_2[O(N)]=0$, a mapping $u$
from S into the $O(N)$ manifold can be extended to a mapping of the solid
ball B into $O(N)$. The WZ term however
is well-defined only modulo a constant.
It was normalized so that if $u$ is  a matrix in the fundamental
representation of $O(N)$ the WZW term is well defined modulo
$WZ\rightarrow WZ+ 2\pi$. The source of the ambiguity is that
$\pi_3[O(N)]\simeq Z$,
namely there are topologically inequivalent ways to
extend $u$ into a mapping from  B into $O(N)$.\par
The bosonic action is invariant under the chiral $O_L(N)\times O_R(N)$
symmetries just as the fermionic action. The associated transformations
are:
$$ u\rightarrow u^{'}=Au \qquad u\rightarrow u^{'}=uB\qquad A,B\subset
O(N).\eqno{\eq}$$ In fact, the invariance holds also for transformation with
$A(x_+)$ and $B(x_-)$, leading to the Kac-Moody algebras of left and
right currents.
 The discussion
of this algebra and the associated currents will be postponed to the
next section.\par The action \mishcb\  is conformaly invariant as well. This
property
was proven by Witten$^{\Wi}$ who originally
showed that if one generalizes \mishcb\  by
taking a coupling $1\over 4\lambda^2$ as a coefficient of the first term
and $k\over 24\pi$ of the WZ term (k integer),
the $\beta$ function
associated with $\lambda$ is given at the one loop level by
$$ \beta \equiv  {d\lambda^2\over dln\Lambda}=-{(N-2)\lambda^2\over
4\pi}[1-({\lambda^2k\over 4\pi})^2], \eqno{\eq}$$
namely eq.\mishcb\  is at a fixed point $\lambda^2={4\pi\over k}$ and hence
exhibits conformal invariance.
By showing that the energy momentum tensor obeys the Virasoro algebra, one
can show that this property is in fact exact.
  \par The bosonic picture for the
theory  of $N$ free massless
Dirac fermions is built from  a boson matrix $g\subset
SU(N)$ and a real boson $\phi$. The bosonized action has now the form
$$ S[g,\phi] = S[g] + \half\int d^2x \partial_\mu \phi\partial^\mu\phi
$$
The construction of the WZ term discussed above applies just as well to  the
$SU(N)$ case. Transformation with respect to global $SU_L(N)\times
SU_R(N)\times U(1)$, $SU_R(N)\times SU_L(N)\times U(1)$
Kac-Moody, and conformal transformations
leave the action invariant.
 One way to prove the equivalence of the theories now,
for $N$ free massless
Dirac fermions and the $k=1$  WZW theory on $U(N)$ group manifold,
is by showing that the
generating functional of the current Green functions of the two
theories are the same. For the fermions we have
$$e^{iW_{\Psi}(A_\mu)} =\int (d\Psi_+d\Psi_-d\bar\Psi_+d\bar\Psi_-)e^{i\int
d^2x
\bar\Psi
 i\ \Dslash\Psi } \eqno{\eq}$$  where here $D_\mu =\partial_\mu +iA_\mu$,
$A_\mu=A_\mu^A({1\over 2}T^A) +A_\mu^{(1)}\times 1$
and $({1\over 2}T^A) \subset SU(N)$.
$W_{\Psi}(A_\mu)$ was calculated by Polyakov and Weigmann
\REF{\Fh}{A.M. Polyakov and P.B. Weigmann, \pl{ 141} (1984)
223.}\refend\
in a regularization scheme   which preserves the global chiral
$SU(N_L)\times SU(N_R)$ symmetry and the local $U(1)$ diagonal symmetry,
\REF{\Fj}{ P. di
Vecchia and P. Rossi, \pl  {140} (1984) 344.}\refend\
\REF{\Fi}{
Y. Frishman, Phys. Lett. {\bf 146B} (1984) 204.
}\refend\
leading
to $$W_{\Psi}(A_\mu)=-S[\squiggle A]-S[\squiggle B]-{1\over 4\pi N}
\int d^2x A_\mu^{(1)}
A^{\mu(1)}
\eqno{\eq}$$
where $\tilde A,\tilde B\subset SU(N)$ are related to the gauge fields
$A^A_\mu$
by  $iA_+^A=(\tilde A^{-1}\partial_+\tilde A)^A, \quad
iA_-^A=(\tilde B^{-1}\partial_-\tilde B)^A $.
\par In the bosonic theory one calculates
$$\eqalign{e^{iW_{B}(A_\mu^A)} =&\int [du]e^{iS[u]+
i\int d^2x (J_-^BA_+^B+J_+^B A_-^B) } \cr
e^{iW_B(A_\mu^{(1)})}=&\int [d\phi]e^{{i\over 2}
\int d^2x[(\partial \phi)^2 +
(J_-A_+^{(1)}+J_+A_-^{(1)})]}\cr}\eqno{\eq}$$
where $J_+^B A_-^B$ and  $J_+A_-^{(1)}$ are the appropriate parts of ${i\over
4\pi}Tr[(g^{-1}\partial_+g)A_-]$, and similarly for the $(-\ +)$ case and with
$A^{(1)}_\pm=Tr (A_\pm)$.
 This functional integrals can be performed exactly,\refmark\Fj  leading
to
$$W_B(A_\mu^A)=-S[\squiggle A]-S[\squiggle B]\qquad W_B(A^{(1)})=
-{1\over 4\pi N}\int d^2x A_\mu^{(1)} A^{\mu(1)}
\eqno{\eq}$$
Thus the bosonic current Green functions are identical to those of the
fermionic theory, the latter  regulated in the way mentioned above.
That $S[u]$ of eqn.\mishcb\ leads to correlation functions for the currents
in the left-right symmetric scheme, can be seen directly\refmark\Fi\
without
performing the functional integral explicitly.

  \section {Non-Abelian bosonization of Dirac fermions with color
and flavor}
\par In his pioneering work on non-abelian bosonization
Witten \refmark\Wi\
 also proposed
a prescription  for bosonizing Majorana fermions which carry both
$N_F$ ``flavors" as well as  $N_C$ ``colors", namely transform under the
group $[ O(N_F)\times O(N_C)]_L \times [ O(N_F)\times
O(N_C)]_R   $. The action for  free
fermions is
$$ S_{\Psi } = {i\over 2}\int d^2x
(\Psi_{-ai}\partial_+ \Psi_{-ai} +
\Psi_{+ai}\partial_- \Psi_{+ai} )\eqno{\eq} $$
where now a=1,....$N_C$, i=1,.....$N_F$ are the color and flavor
indices respectively. The equivalent bosonic action is
$$ \squiggle S[g,h]={1\over2} N_cS[g]+{1\over2}N_F S[h]\eqn{\mishce}$$
The bosonic fields $g$ and $h$ take their values in $O(N_F)$ and $ O(N_C)$
respectively and  $S[u]$ is the WZW action given in \mishcb\ .

The bosonization dictionary for the currents was shown to be:
$$ J_{+ij} =:\Psi_{+ai} \Psi_{+aj}:= {iN_C\over 2\pi}(g^{-1}\partial_+g)_{ij}
\qquad
J_{-ij} =:\Psi_{-ai} \Psi_{-aj}:= {iN_C\over 2\pi}(g\partial_-g^{-1})_{ij}
\eqn{\mishcf}$$
$$ J_{+ab} =:\Psi_{+ai} \Psi_{+bi}:={iN_F\over 2\pi}(h^{-1}\partial_+h)_{ab}
\qquad
J_{-ab} =:\Psi_{-ai} \Psi_{-bi}:= {iN_F\over 2\pi}(h\partial_-h^{-1})_{ab}
\eqn{\mishcg} $$
where :   : stands for normal ordering with respect to fermion creation
and anihilation operators.
As for the bosonic expressions for the currents, regularization is obtained by
subtracting the appropriate singular parts.\REF{\KZ}{
V. G. Knizhnik and A. B. Zamolodchikov, \np {247} (1984) 83.}

This procedure was already used in connection with the Thirring model in ref.
[\DFZ], and for the case at hand in ref.[\KZ]. It is actually much older,
as in Schwinger's ``point splitting"  technique\REF{\Schw}{J. Schwinger,
\prl {3} (1959) 269.}\refend,
and in general from Wilson's operator product
expansion.\REF{\Wil}{K. Wilson, \PR {179} (1969) 1499.}\refend
 \par In terms of the complex coordinates $z=\xi_1+i\xi_2,
\quad\bar z=\xi_1-i\xi_2$  (where $\xi_1$ and
$\xi_2$ are complex coordinates spanning $C^2$,
and the Euclidian plane $(\xi_1\rightarrow x,\xi_2\rightarrow -t)$
and Minkowski
space-time $(\xi_1\rightarrow x, \xi_2\rightarrow -it)$
can be obtained as appropriate real sections), one can express the
currents as
 $$ J(z)_{ij}\equiv\pi J_{-ij}= {iN_C\over 2}(g\partial_zg^{-1})_{ij} \qquad
 \bar J(\bar z){ij}\equiv \pi J_{+ij}
= {iN_C\over2}(g^{-1}\partial_{\bar z}g)_{ij} \eqno{\eq} $$
and similarly for the colored currents.
\par In a complete analogy the theory of $N_F\times N_C$ Dirac
fermions can be expressed in terms of the bosonic fields
$g$, $h$, $e^{-i\sqrt{4\pi\over N_FN_C}\phi}$
 now in $SU(N_F), SU(N_C) $ and U(1)
group manifolds respectively.\REFS
{\Fk}{D. Gonzales and A.N. Redlich, Nucl. Phys.
{\bf B256} (1985) 621.}
\REFSCON{\Fl}{I. Affleck, Nucl. Phys. {\bf B265}
[FS15] (1986) 448. }
\refsend\
The corresponding  action is now:
$$ S[g,h] = N_CS[g]+N_FS[h] +{1\over 2}\int d^2x
 \partial_{\mu}\phi \partial^{\mu} \phi \eqn{\mishck}$$
This action is derived simply by substituting  $ ghe^{-i\sqrt{4\pi\over
N_CN_F}\phi}$ instead of $u$ in \mishcb .

 As for the  equivalence between the bosonic and fermionic
theories, we note that in  both theories the
commutators of the various currents have the same current algebra,
and the
energy-momentum tensor is the same when expressed in terms of the
currents. But the situation changes when mass term are introduced ( see next
section 3.4).
The bosonization rules for the color and flavor
currents are obtained from  \mishcf\ and \mishcg\
by replacing the Weyl-Majorana spinors with Weyl ones, and in addition
we have  the U(1) current
$$\eqalign{
J^{(1)}(z)\equiv \sqrt{\pi} J_-^{(1)}=&
:\Psi^\dagger _{-ai} \Psi_{-ai}:= \sqrt{N_FN_C\over \pi}\partial_-\phi\cr
 \bar J^{(1)}(\bar z)\equiv \sqrt{\pi} J_+^{(1)}=&
:\Psi^\dagger _{+ai} \Psi_{+ai}:
= \sqrt{N_FN_C\over \pi }\partial _+ \phi\cr}\eqno{\eq} $$
\par The Kac-Moody
algebras are given by:
$$ [J^A_n,J^B_m] =if^{ABC}J^C_{n+m}+{i\over 2}kn\delta ^{AB}\delta_
{n+m,0} \eqno{\eq}$$
where $J^A =Tr(T^AJ)  ,\ T^A$  the matrices of $SU(N_C)$,
$k=N_F$  for the
colored currents and $J(z)$ is expanded in a Laurent series as
$ J(z) =\sum z^{-n-1}J_n $. Similar expression will apply for
the flavor currents with $T^I$ the matrices of $SU(N_F)$, and the
central charge  $k= N_C$ instead of $ N_F$. The commutation
relation for$ \bar J(\bar z) $ will  have the same form.
\par Generalizing the case of $SU(N)\times U(1)$\refmark\FD\
to our case,
the Sugawara form\refmark{\Sug}
 for the energy
momentum tensor of the WZW action  is given by:
$$\eqalign{T(z) =&{1\over 2\kappa_C}\sum_A :J^A(z)J^A(z):
+ {1\over 2\kappa_F}\sum_I :J^I(z)J^I(z):\cr
&+ {1\over 2\kappa} :J^{(1)}(z)J^{(1)}(z):\cr} \eqno{\eq}$$
where the dots denote normal ordering   with respect to $n$
($n>0$ meaning anihilation).
 The $\kappa$'s are constants yet to be determined. In terms of the
Kac-Moody generators this can be written as
$$\eqalign{L_n =&
{1\over 2\kappa_C}\sum_{m=-\infty}^{\infty} :J^A_mJ^A_{n-m}: +
{1\over 2\kappa_F}\sum_{m=-\infty}^{\infty} :J^I_mJ^I_{n-m}:\cr
&+{1\over 2\kappa}\sum_{m=-\infty}^{\infty} :J^{(1)}_mJ^{(1)}_{n-m}:\cr}
\eqno{\eq}$$
Now, by applying the last expression on any primary
field $\phi_l$  we can get a
set of infinitely many ``null vectors"of the form
$$\chi_l^n= [L_n-
{1\over 2\kappa_C}\sum_{m=n}^0 :J^A_mJ^A_{n-m}: -
{1\over 2\kappa_F}\sum_{m=n}^0 :J^I_mJ^I_{n-m}:
-{1\over 2\kappa}\sum_{m=n}^0 :J^{(1)}_mJ^{(1)}_{n-m}:
]\phi_l\eqno{\eq}$$
for any $n\leq 0$ (for $n>0$ holds immediatelly).
Since each of these vectors must certainly be a
primary field, $L_m\chi^n =J^A_m\chi^n =J^I_m\chi^n =J_m\chi
^n =O$  for $m>0$ holds. This
leads to expressions for the various $\kappa$, for
the central charge
$c$ of the Virasoro Algebra  and for
the dimensions of the primary fields $\Delta_l=\Delta_{l+}+\Delta_{l-}$,
in terms of $N_C,N_F$ and the group properties of the
primary fields:
$$ \kappa_C =\kappa_F ={1\over 2}(N_C+N_F),
\qquad \kappa =N_FN_C \eqn\mishccl$$
$$c
={N_C(N_F^2-1)\over(N_C+N_F)}
+{N_F(N_C^2-1)\over(N_C+N_F)}+1=N_FN_C.\eqno{\eq}$$   \par
$$  \Delta_{l\pm}= {(c_{l\pm}^2)^F\over{(N_F+ N_C)}}
+ {(c_{l\pm}^2)^C\over{(N_F+ N_C)}}+
  {(c_{l\pm}^2)^{(1)}\over{N_CN_F}} \eqno{\eq}$$
where $(c_{l\pm}^2)^C$ is the eigenvalue of the $SU_{R,L}(N_C)$
second Casimir operator in the
representation of the primary field $\phi_l$, namely
$(\half T^A)(\half T^A)
=(c_l^2)^CI$, and similarly for the flavor group.
\REF\FSC{ Y. Frishman and J. Sonnenschein, \np {301} (1988) 346.}
\REF\SoCh{J. Sonnenschein, \np {309} (1988) 752.}
In the cases of $SU(N_C)$  and $SU(N_F)$ the discussion   applies to
$\Delta_{l+}$ or $\Delta_{l-}$ separately, with $C^2_{l+}$ and $C^2_{l-}$
respectively. For the $U(1)$ factor we discuss only  the total dimension, as we
do not wish to enter here    into the issue of chiral bosons ([\FSC,\SoCh] and
references therein). Note that the expressions for $\kappa_F$ and $\kappa_C$
of
eqn. \mishccl\ are an immediate generalization of ref.[\FD]  in which the group
$SU(N)$ was discussed with central term equal to one. There the factor was
$N+1$, the $N$ being the second Casimir of the adjoint representation, and
the $1$ being the central term.
 \par The
equivalence of the bosonic and fermionic Hilbert spaces
was demonstrated by showing that the two theories have the same
current algebra (Kac-Moody algebra), and that  the energy-momentum tensor
can be constructed from the currents in a Sugawara form.
 Goddard et al \REF{\Fn}{
P. Goddard, W. Nahm and D. Olive, \pl        {160} (1985) 111.}\refend\
showed that a necessary  and sufficient condition for such
a construction of the fermionic $T(z) $, in a theory with a
symmetry group G, is the existence of a larger group
$G\subset G'$  such that $G'/G $
is a symmetric space with the fermions transforming under G just as the
the tangent space to $G'/G $ does.
Based on this theorem they found all the fermionic
theories for which an equivalent  WZW bosonic action can be constructed.
The cases stated above fit in this category.
Note in passing that this does not hold for
cases where the symmetry group includes more non abelian group factors,
like for example $SU(N_A)\times SU(N_F)\times SU(N_C)\times U(1)$.
\REF\BFSc{T. Banks, Y. Frishman and A. Schwimmer, unpublished (1984).
}\refend
\par The prescription eqn. \mishck\
described above , for the bosonic action
that is equivalent to that of colored and flavored Dirac fermions, is by
no means unique. In fact it will be shown that this prescription will
turn out to be inconvenient once mass terms are introduced. Another
scheme,  based on the WZW theory of $U(N_FN_C)$ will be recommended.
\sect{ The bosonization of a mass bilinear of Dirac fermions}
\par A further bosonization rule
has to be invoked  for the mass bilinear. For a theory with  a U(N)
symmetry group the rule is the following:
$$ \Psi_+^{\dagger l}\Psi_{-j}
=\tilde c  \mu  N_{\mu} g^l_j e^{-i\sqrt{4\pi \over N
}\phi}\eqn{\mishcl} $$
where $N_{\mu}$ denotes
normal ordering at mass scale $\mu$ .
The fermion mass term $m_q\bar \Psi^i \Psi_i$ is therefore
$$ m^{'2} N_{\mu}\int d^2x Tr(g+g^\dagger)\eqno{\eq}$$where
$m^{'2}=m_q\tilde c\mu$,
$m_q$ is the quark mass, and $c$ is the same
constant  as in eqn. (2.1).
It is straightforward to show that
the above bosonic operator transforms correctly  under the
$U(N)_L\times U(N)_R $ chiral transformations. On top of that it has
the correct total dimension
$$ \Delta= \Delta_g +\Delta_{\phi} = ({N-1\over N}+{1\over
N})=1 \eqno{\eq} $$
where $ \Delta_g ={N-1\over N}$ and $ \Delta _\phi ={1\over N}$ are the
dimensions associated with the SU(N) and U(1) group factors respectively.
Moreover it was explicitly shown that the four point function
$$ G(z_i,\bar z_i) = <g(z_1,\bar z_1)g^{-1}(z_2,\bar z_2)
g^{-1}(z_3,\bar z_3)g(z_4,\bar z_4)> \eqno{\eq} $$
is given by\refmark\KZ  :

$$ G(z_i,\bar z_i) = [(z_1- z_4)(z_2- z_3)(\bar z_1-\bar z_4)
(\bar z_2-\bar z_3)]^{-\Delta_g} G(x,\bar x)  \eqno{\eq}$$
$G(x,\bar x)$ is the following function of the harmonic quotients
\bk$ x={(z_1-z_2)(z_3-z_4)\over (z_1-z_4)(z_3-z_2)}\qquad $ and $
\bar x={(\bar z_1-\bar z_2)(\bar z_3-\bar z_4)
\over (\bar z_1-\bar z_4)(\bar z_3-\bar z_2)}$,
$$ G(x,\bar x)=[x\bar x (1-x)(1-\bar x)]^{1\over N}\times
[I_1{1\over x}+I_2{1\over 1-x}][\bar I_1{1\over \bar x}+\bar I_2
{1\over 1-\bar x}], \eqn\mishcr$$
where $I_1,I_2,\bar I_1,\bar I_2$ are group
invariant factors.
This result for the correlation function, combined with the $U(1)$ part
gives an expression identical to that for the
fermionic bilinears.
Moreover the  result can be generalized to an n-point function.
The method of computing four point functions in conformally invariant theories
was suggested already  in ref. [\FD].

 \sect {Bosonization of mass bilinears  in the   product scheme}
\par A natural question here is how to generalize the rule \mishcl\  to
theories given by \mishce\  and its analog for the case of
$SU(N_F)\times SU(N_C)\times U(1)$
given in \mishck .  We call the latter the {\bf product scheme}.
It was argued that
the  bosonization rule for the latter case
is \refmark\Fk
$$ \Psi_+^{\dagger ai}\Psi_{-bj} =\tilde c\mu N_{\mu} g^i_j h^a_b
e^{-i\sqrt{4\pi \over N_FN_C}
\phi}\eqn{\mishcs} $$
Consequently, the bosonic form of
the fermion mass term $m_q\bar \Psi^{ia} \Psi_{ia}$
is therefore
$$ m^{'2} N_{\mu}\int d^2x(TrgTrh+Trh^\dagger Trg^\dagger)
e^{-i\sqrt{4\pi \over N_FN_C}\phi}\eqn\mishaha$$
with $m^{'2}=m_q\tilde c\mu$.
Once again the bosonic operator \mishcs\
has the correct chiral transformations and the proper dimension:
$$ \Delta= \Delta_g +\Delta_h +\Delta_{\phi} =
{N_F^2-1\over{N_F(N_F+ N_C)}} + {N_C^2-1\over{N_C(N_C +N_F)}}+
{1\over{N_CN_F}}=1 \eqno{\eq} $$
Unfortunately, the explicit calculation of the
four point function\refmark\KZ
\ reveals a discrepancy between the fermionic and bosonic results.
This can actually be understood directly.
Since $g$
and $h$ are fields defined on entirely independent group manifolds, then
(ignoring for a moment the U(1) factor) the analog of (3.26)
can be written as :
$$  <g(z_1,\bar z_1)g^{-1}(z_2,\bar z_2)
g^{-1}(z_3,\bar z_3)g(z_4,\bar z_4)> <h(z_1,\bar z_1)h^{-1}(z_2,\bar z_2)
h^{-1}(z_3,\bar z_3)h(z_4,\bar z_4)> \eqno{\eq}$$
This expression  differs from the
corresponding
fermionic Green's function in at least two aspects: (i) It includes
independent ``contractions" for the $g$ and $h$ factors, whereas in the
fermionic correlation function the flavor and color contractions
are correlated.
(ii) The result \mishcr\
 is true only for a bosonic field associated with
Kac-Moody central charge $k=1$ . For $g$ and h, however, the central charges
are $N_C$ and $N_F$ respectively. For such cases the expression for the
Green's function is much more complicated (expressed in terms of
hyper-geometric functions) and does not resemble  the case
of free fermions.
\sect {Bosonization using the  $U(N_F\times N_C)$ WZW action}
\par
It is clear from the previous discussion
that the bosonization prescription for our case
needs an alteration. Apriori there can be two  ways out: modifying
the rule for the bosonization of the mass bilinear
or  using a different
bosonic theory altogether.
As for the first approach, eq. \mishcs\  preserves  the  proper chiral
transformation laws under the product group
$SU(N_F)\times SU(N_C)\times U(1)$
as well as the correct dimension,  and therefore the
number of possible modifications is very limited. For example one might
think of multiplying the expression in eq.
\mishcs\  by an operator which is a chiral singlet under the above group,
with zero dimension. We do not know of
such a modification. It may possibly be non-local.
Therefore we are going to try a different bosonic theory than eq. \mishck\ .
The symmetry of the free fermionic theory can  actually be taken as
 $U_L(N_F\times N_C)\times U_R(N_F\times N_C) $ rather than
$[ SU(N_F)\times SU(N_C)\times U(1)]_L \times
[ SU(N_F)\times SU(N_C)\times U(1)]_R $.
 The natural bosonic action is
hence  a WZW theory of $u\subset U(N_FN_C) $ and with  $k=1$ .
The action will be given by eqn. (3.2). The currents are now
 $$ J(z)_{\alpha\beta}= {i\over 2}(u\partial_zu^{-1})_{\alpha\beta}
\qquad   \bar J(\bar z)_{\alpha\beta}= {i\over
2}(u^{-1}\partial_{\bar z}u)_{\alpha\beta}\eqn\mishcc$$
with $\alpha,\beta$ running from 1  to $N_F\times N_C$. The formulas
(3.11-3.12)
can be obtained from \mishcc\  by appropriate traces, over color for (3.11)
and over flavor for (3.12). The mass bilinear   is now
$$ \Psi^{\dagger}_{+\alpha}\Psi_{-\beta } =
\tilde c\mu  N_{\mu}   u_{\alpha\beta}
\eqn{\mishccl} $$
which is as (3.23) but with the $U(1)$ term  absorbed in $u$.

Clearly the requirement for Sugawara construction of T, for proper
chiral transformations of all the operators and for a correct
dimension for the mass bilinear are fulfilled.  Since now the flavor and
color degrees of freedom are attached to the same bosonic field, the
previous ``contraction problem" in the n point functions
is automatically resolved. Moreover as  stated
above the four-point function   and in fact any Green's function
will now reproduce the results of the
fermionic calculation  .\par
The currents constructed from $u$ obey the Kac-Moody algebra with $k=1$.
The color currents, for instance, are $J^A=Tr(T^A J)$,  where $T^A$
are expressed as
$(N_CN_F)\times (N_CN_F)$ matrices
defined by  $\lambda ^A \otimes 1$,
with $\lambda ^A$ the
Gell-Mann matrices in color space and 1 stands for a unit $N_F\times N_F$
matrix.
The central charge is
$k=N_F$. The same arguments will apply for the flavor currents,
now with $k=N_C$.
The central
charge for the $U(1)$ current is $N_CN_F$.\par
To see the difference between the present theory and the previous one let
us express $u$ in terms of $(N_FN_C)\times (N_FN_C)$ matrices
$\tilde g ,\tilde h $ and $\tilde l $
in $SU(N_F) ,SU(N_C) $ and the coset-space
$ SU(N_F \times
N_C ) /\{ SU(N_F)\times SU(N_C)\times U(1)\} $ respectively, through
$ u=\tilde g \tilde h\tilde l e^{-i\sqrt{4\pi\over N_C N_F}\phi}
$. Using
the formula for expressing an action of the form $S[AgB^{-1}]$
\refmark\Fh\
we get:
$$\eqalign{S[ u] &= S[\tilde g \tilde h
\tilde l] +
{1\over 2} \int d^2x\partial_\mu\phi\partial^\mu\phi\cr
S[\tilde g  \tilde h \tilde l]&=
S[\tilde g ] + S[\tilde  l] + S[\tilde h ] +{1\over 2\pi}
\int d^2x
Tr(\tilde g^\dagger \partial_+\tilde g \tilde l\partial_-\tilde l^\dagger
+ \tilde h^\dagger \partial_+\tilde h
\tilde l \partial_-\tilde l^\dagger)\cr}
\eqno{\eq})$$
We can now choose $\tilde
l=l$ so that $l\partial_-l^\dagger $ will be spanned by the generators
that are only in
$ SU(N_F \times
N_C ) /\{ SU(N_F)\times SU(N_C)\times U(1)\} $.
This can be achieved by taking $ \tilde u=\tilde g \tilde h\tilde l $
  ( namely $u$ but without the $U(1)$ part),
and then taking for $\tilde h= h \otimes 1 $
a solution of the equation
$\partial_- h h^\dagger={1\over N_F}
Tr_F[(\partial_-\tilde u) \tilde u^\dagger]$
, and  similarly
for $g$ with ${1\over N_C} Tr_C$ .
These are also the conditions that the flavor currents be expressed in
terms of $\tilde g$ and the  color currents in terms of $\tilde
h$.
For this choice, the mixed  term in the action (3.35) , the one involving
products of $\tilde l's $ with $\tilde g's$
or  $\tilde  h's$ is zero, and so
the new action is
 $$S[u] = N_C S[ g ] + N_F S[h] +
{1\over2} \int d^2x\partial_\mu\phi\partial^\mu\phi
+ S[l ]\eqno{\eq})$$
Note that $l$ is still an $SU(N_CN_F)$ matrix while $g$ and $h$
are expressed now as $SU(N_F)$ and $SU(N_C)$ matrices respectively,
but the
matrix $l$ involves only products of clolor and flavor matrices
(not any of them separatelly).
\par \endpage
\chap {Bosonized  Massive Multiflavor QCD$_2$}
In the last chapter we derived the bosonization rules for
 massive fermions which
transform under  flavor and  color symmetry groups. Here we develop the
bosonic version of multiflavor massive $QCD_2$. Applying the strong coupling
limit we deduce the low energy effective action.\refmark{\DFS,\FS}
In the product scheme\refmark\DFS\ this effective action comes out to
depend on color singlet variables only, the $g$'s. It presents a
derivation for a two dimensional ``Skyrme model",\refmark\Skyrme
 which in four
dimensions was not derived directly,\refmark\SM
but argued on general grounds.

\sect{ The ``Hybrid" approach}
Before proceeding to gauge the
color symmetry group of the colored flavored
WZW model, we describe briefly another approach introduced by Gepner
\REF{\Gepner}{D. Gepner,
\np {252} (1985) 481.}\refend
 in which the flavor
sector appears in the form of a
WZW model but for the color degrees of freedom
the gauged abelian bosonization is invoked. Recall the multiflavor $QCD_2$ in
the abelian formulation (see section 2.6).
In the gauge  eqn.(2.32) one can use
the Gauss law eqn. (2.33)
to express the gauge fields in terms of the appropriate
fermionic bilinears,
which translate into bosonic group elements as follows:
 $$\partial_1 e_a =\sqrt{\pi}\Sigma_i\Psi^{\dagger ia}\Psi_{ia} = {i\over
\sqrt{\pi}}\partial_1Tr log g_a\eqn \mishddr$$ where $g_a\in U(N_F)$ is one out
of $N_C$ such matrices, and $e_a$ connects to the original electric
fields  by $2\sqrt{\pi}E^a_a =(e_a-{1\over
N_C}\Sigma_k e_k)$. One can now use again eqn. (2.33) again
to also
 express $A^b_a$ for $a\ne b$ in terms of fermion densities.
Inserting these into the $QCD_2$ Hamiltonian given in eqn. (2.31)
Gepner gets\refmark\Gepner
$$\eqalign{{\cal H} =&{\cal H}^0 +{\cal H}^I\cr
{\cal H}^I =& -\sum_{a,b}{(e_c)^2\over
32\pi^2N_C}\(Trlog(g_ag_b^{-1})\)^2
-\sum_{a,b}
\pi\mu^2{Tr(g_ag_b^{-1})\over
Trlog(g_ag_b^{-1})}+\sum_a mc\mu\sqrt{N_F}Tr(g_a)\cr}\eqn{\mishHam}$$
${\cal H}^0$ includes the fermion kinetic term.
For $N_F=2$ the potential is free from singularities, for $N_F\ge 3$ it is
not. In ref. [\Gepner]\
the case of $N_F=2$ was analyzed thouroghly
deducing the corresponding low lying baryonic spectrum. In the present review
the full WZW is emphazied,
so we refer the reader to [\Gepner] for  details in
the framework of the ``Hybrid " approach.

 \section {Gauging the WZW action}
 \par  Coming back to the full WZW
 description   of colored and flavored  fermions,  the
next step toward the bosonized version of QCD$_2$ is obviously the introduction
of the color interaction.
This is achieved by  gauging the  the vector subgroup
$SU_V(N_C)$ of $SU_L(N_C)\times SU_R (N_C)$.
There are various methods to gauge the model.
Here we present two of them :
  a trial and error method,
\REF{\Fo}{E. Witten, \np{223} (1983) 422.}\refend\
and gauging via
 covariantizing the current. Those methods
are applicable also in  the $U(N_FN_C)$ bosonization scheme.

(4.2a) Trial and error Noether method

The WZW action on $SU(N_C)$ group manifold
is, as stated above, invariant under the
global vector transformation $ h\rightarrow UhU^{-1}$ where $U\subset
SU(N_C)$. Now we want to vary $h$ with respect to the associated local
infinitesimal transformation $U=1+i\epsilon(x)= 1+iT^A\epsilon^A(x)$
$$\delta_\epsilon h=i[\epsilon,h]\qquad
\delta_\epsilon h^{-1}=i[\epsilon,h^{-1}]\eqno{\eq}$$
The variation of the action $S^{(0)}[h] \equiv S[h]$  under such a
transformation is obviously given by:
$$ \delta_\epsilon S^{(0)}[h]
=-\int d^2x Tr(\partial_\mu \epsilon J^{\mu}) \eqno{\eq}$$
where the Neother vector current is given by:
$$ J_\mu  ={i\over4\pi}\{[h^\dagger \partial_\mu h + h\partial_\mu
h^\dagger]-
\varepsilon_{\mu \nu }[h^\dagger\partial^\nu h -h\partial^\nu h
^\dagger ] \} \eqn{\mishbc}$$
We introduce now the first correction term $S^{(1)}$ given by
$$S^{(1)} =\int d^2x Tr(A_\mu J^{\mu}) \qquad
\delta_\epsilon S^{(1)}[h]
=-\int d^2x Tr[\partial_\mu \epsilon(J^{\mu}+J^{'\mu})] \eqn{\mishba}$$
The second part of eqn. \mishba\
 is the variation of $S^{(1)}$ which is derived
 using the infinitesimal variation of the gauge field $\delta A_\mu=
-D_\mu \epsilon =-(\partial_\mu \epsilon + i[A_\mu,\epsilon]) $.
$J^{'\mu}$ is found to be:
$$ J^{'}_\mu  ={-1\over4\pi}\{[h^\dagger A_\mu h + hA_\mu
h^\dagger -2A_\mu]-
\varepsilon_{\mu \nu }[h^\dagger A^\nu h -hA^\nu h
^\dagger ] \} \eqno{\eq}$$
The second iteration will be by adding $S^{(2)}$ where now $J^{'\mu}$ is
replacing $J^{\mu}$.
$$S^{(2)} =\int d^2x Tr(A_\mu J^{'\mu}) \qquad
\delta_\epsilon S^{(2)}[h]
= -2\int d^2x Tr(\partial_\mu \epsilon J^{'\mu}).\eqno{\eq}$$
It is therefore obvious that
$$ \delta_\epsilon [S^{(0)} +S^{(1)}-\half S^{(2)}]=0
\eqno{\eq}$$
Hence $S[h,A_\mu]\equiv [S^{(0)} +S^{(1)}-\half S^{(2)}]$
is given by
$$\eqalign{S[h,A_\mu] &=
{1\over8\pi}\int d^2xTr(D_\mu h D^\mu h^\dagger)
+{1\over12\pi}\int_B d^3y\varepsilon^{ijk}Tr(h^\dagger\partial_ih)
(h^\dagger\partial_jh)(h^\dagger\partial_kh)\cr &\qquad -{1\over4\pi}
\int d^2x\varepsilon_{\mu\nu}
Tr[iA^\mu(h^\dagger\partial^\nu h-h\partial^\nu h^\dagger+ih
^\dagger A^\nu h)] \cr} \eqn{\mishbd}$$
which can also be written in light
cone coordinates
$$\eqalign{ S[h,A_+,A_-]
&=S[h]+{i\over2\pi}
\int d^2xTr(A_+h\partial_-h^\dagger+A_-h^\dagger\partial_+h)\cr
&-{1\over2\pi}\int d^2xTr(A_+hA_-h^\dagger-A_-A_+)\cdot\cr}
\eqn{\mishbb}$$

(4.2b) Gauging via covariantization of the Neother current

In four space-time dimensions the current, in terms of bosonic matrices,
involves up to third power\refmark\SM\
 in the gauge potentials.
 In D space-time dimensions the bare current
 will contain
 (D-1) derivatives and is gauged by replacing the ordinary
derivatives with covariant derivatives and by adding terms which contain
products of  $F_{\mu \nu}$ with powers of $h$ and $h^\dagger$ and covariant
derivatives $D_\mu h$  and $D_\mu h^\dagger$. In two dimensions, however, there
is no room for such  terms in the gauge covariant  current, as
these terms
involve $\epsilon_{\mu_1.....\mu_D}$ in D dimensions with one free
index and the others contracted with $F_{\mu \nu}$'s and $D_\mu$'s
, and in two dimensions they
cannot be constructed.
Therefore the covariantized current is given by:
$$ J_\mu(h,A_{\mu})  ={i\over4\pi}\{[h^\dagger D_\mu h + h D_\mu
h^\dagger]-
\varepsilon_{\mu \nu }[h^\dagger D^\nu h -h D^\nu h
^\dagger ] \} \eqno{\eq}$$
Knowing the current we deduce the action via $J_\mu =
{\partial S\over \partial A_\mu} $ getting straightforwardly eq.
\mishbb\ .
\par
Finally, we combine the gauged WZW action of the color group manifold,
the WZW of the flavor group manifold and the action term for the gauge
fields, to get the bosonic form of the
action of massless QCD$_2$. The well known fermionic form of the action
is (a mass term will be added later),
$$S_F[\Psi,A_\mu]=\int d^2x\{-{1\over2e_c^2}Tr(F_{\mu\nu}
F^{\mu\nu})-\bar \Psi^{ai} {[ (i\dslash +\Aslash) \psi_i ]}_a\}
\eqno{\eq}$$
where  $e_c$ is the coupling constant to the color potentials (note it has
mass dimensions in 1+1 space-time), and
 $$ F_{\mu\nu}=\partial_\mu A_\nu -\partial_\nu A_\mu
+i[A_\mu ,A_\nu ] $$
 The bosonized action is
$$\eqalign{ S[g,h,A_+,A_-] &= N_cS[g]+N_FS[h]\cr
&+ {N_F\over 2\pi}\int d^2xTr [ i(A_+h\partial_-h^\dagger
+A_-h^\dagger\partial_+h)-(A_+hA_-h^\dagger-A_-A_+)]\cr
& -{1\over 2e^2_c}\int d^2 xTr
F_{\mu\nu} F^{\mu\nu} \cdot \cr} \eqno{\eq}$$
\sect{ The strong coupling limit}
So far we wrote the bosonic version of massless $QCD_2$. It  turns out that
the mass term plays an important role in the determination of classical soliton
solutions  in 1+1 space-time dimensions. It is therefore required to switch on
this term before deducing the low energy effective action. As was explained in
the last chapter this we know how to do rigorously   only in the
 scheme of $U(N_FN_C)$. Nevertheless, we want first to integrate over the
gauge fields and high energy modes in the product scheme  eqn. (4.13). In the
next section we repeat the derivation in the other scheme.
It will turn out that the product scheme can be used for the low  mass states
in the strong coupling limit.
 \par We first have to choose a gauge. We choose a light-cone gauge  $A_-=0$.
This has two advantages, namely  no Fadeev-Popov  ghosts and also no
interaction terms  quadratic  in $A$. The latter fact makes it immediate to
integrate out the gauge potentials.    Define  $ H(x) $ by $ \partial_- H
=i h\partial_- h^\dagger $  with $ H (-\infty ,x_-) =0$. Then also
$ H (\infty ,x_-) =0$ for colorless states.
 Now we add the mass term as in eqn. (3.30) , absorbing the $U(1)$
factor in $g$, and then integrate out $A_+$\refmark\DFS\
to obtain $$\eqalign{
S[g,h] &= N_cS[g]+N_FS[h]\cr &-\half({N_F e_c\over2\pi})^2\int
d^2x Tr( H^2) + m^{'2} N_{\mu}\int d^2x(TrgTrh+Trh^\dagger Trg^\dagger)
\cr}\eqn\mishdl$$ where $Tr h$  is over color, $Tr g$ over flavor, etc.
In the strong coupling limit $e_c/m_q\rightarrow \infty$ the low energy
effective action reads:
$$ S[g] = N_cS[g]
 + m^{2} N_{m}\int d^2x(Trg+
Trg^\dagger) \eqn\mishdl$$
The determination of the  parameter $m$  obtained after appropriate normal
ordering, will be explained in the next section.
Note that the analog of our strong coupling to the case of 3+1 space time,
would be that of light current quarks compared to the $QCD$ scale
$\Lambda_{QCD}$.
 \sect{ Multiflavor $QCD_2$ using the $U(N_F\times N_C)$ Scheme}
 \par Let us  now repeat the gauging of the
 $SU_V(N_C)$
 subgroup
in the framework of the $U(N_F\times N_C)$
bosonization procedure.
\par Using the gauging prescription discussed in section 4.2      we get
first the action where the whole $SU(N_CN_F)$ is gauged, namely
$$\eqalign{ S[u,A_+,A_-]
&=S[u]+{i\over2\pi}
\int d^2xTr(A_+u\partial_-u^\dagger+A_-u^\dagger\partial_+u)\cr
&-{1\over2\pi}\int d^2xTr(A_+uA_-u^\dagger-A_-A_+)
+ m^{'2} N_{\tilde m}\int d^2x Tr(u+u^\dagger)\cr}
\eqno{\eq}$$
where we have also added a mass term with
$m^{'2} =m_qc\tilde m $.
Now since we are interested in gauging only the $SU(N_C)$ subgroup
of $U(N_FN_C), $ we take $A_\mu$
to be   related to the generator $T^D \subset SU(N_C) $
via $A_{\mu}=e_c A^D_{\mu} T^D $.
We then add to this action the kinetic term for the gauge fields
$-{1\over2e_c^{'2}}\int d^2x Tr(F_{\mu \nu}F^{\mu \nu})$.
 $e_c ^{'}$ has the appropriate
form so that after
tracing in flavor space the color gauge coupling is obtained,
namely $e_c^{'}={\sqrt N_F}e_c $.
The  resulting action is invariant under:
$$ u\rightarrow V(x)uV^{-1}(x) \qquad A_{\mu} \rightarrow
V(x)(A_{\mu}-i\partial_{\mu} )V^{-1}(x)
\qquad V(x) \subset SU_V(N_C)\eqno
{\eq} $$
$$ u\rightarrow WuW^{-1} \qquad W\subset U(N_F) \eqno{\eq} $$
The symmetry group is now $SU_V(N_C)\times U(N_F)$,
just as for the gauged fermionic theory.
We choose the gauge $A_-=0$, so now the action takes the form:

$$\eqalign{ S[u,A_+] &=S[u] +{1\over e_c^{'2}}\int
d^2x Tr(\partial_-A_+)^2
+{i\over 2\pi } \int d^2x Tr (A_+u\partial_-u^\dagger ) \cr
&+ m^{'2} N_{\tilde m} \int d^2x Tr(u+u^\dagger)\cr}\eqno{\eq} $$
Upon the decomposition
$ u=\tilde g \tilde h l e^{-i\sqrt{4\pi\over N_C N_F}\phi}$ we see
that the current that couples to $A_+$ is
$ \tilde h\partial_-\tilde h^\dagger $.
In terms of $u$ it is the color projection $(u\partial_-u^\dagger)_C=
{1\over N_F}
Tr_F\(u\partial_-u^\dagger-{1\over N_C}Tr_C u\partial_-u^\dagger\)$.
Thus the coupling of the current to the gauge field
${i\over 2\pi}\int d^2x Tr (A_+\tilde h\partial_-\tilde h^\dagger)$.
To proceed we define $\tilde H(x) $ by
$ \partial_-\tilde H =i\tilde h\partial_-\tilde h^\dagger $.
We take the boundary conditions to be $\tilde H (-\infty ,x_-) =0$
and then we integrate out $A_+$ obtaining
$$\eqalign{ \tilde S[u] &=S[u] -({e_c\over4\pi})^2
 N_F\int
d^2x Tr(\tilde H^2)\cr
&+ m^{'2} N_{\tilde m} \int d^2x Tr(u+u^\dagger)\cr}\eqno{\eq} $$
In the strong coupling limit $ {e_c\over m_q} \rightarrow \infty $
the fields in $\tilde h$
which contribute to $\tilde H $ will become infinitely heavy. The
sector $\tilde gl \subset
{SU(N_FN_C)\over SU(N_C)} $,
however, will not acquire mass from the gauge interaction term.
Since we are interested only in the light particles we can in the strong
coupling limit ignore the heavy fields, if we first
normal order the heavy fields at the mass scale
$\tilde \mu ={e_c\sqrt{N_F} \over \sqrt{2\pi}}$.
Using the relation, for  a given operator $O$,
$$ ({\tilde \mu \over \tilde m})^\Delta N_{\tilde \mu} O =
N_{\tilde m}O \eqno{\eq}$$
to perform the change in the  scale of
normal ordering,
and then substituting  $h^a_b =\delta^a_b,$ we get for the low energy
effective action:
$$\eqalign{ S_{eff}[u]
&=S[\tilde g]+S[l]+
{1\over 2} \int d^2x\partial_\mu\phi\partial^\mu\phi\cr
&+ cm_q\tilde \mu
N_{\tilde \mu} \int d^2x Tr(e^{-i\sqrt{4\pi\over N_C N_F}\phi}
\tilde gl + e^{+i\sqrt{4\pi\over N_C N_F}\phi}
\l^\dagger \tilde g^\dagger )\cr}. \eqno{\eq}$$
We can now replace the two mass scales $m_q$ and $\tilde\mu$
by a single scale by
normal ordering at a certain $m$ so the final form of the effective
action becomes
$$\eqalign{ S_{eff}[u]
&=S[\tilde g]+S[l]+
{1\over 2} \int d^2x\partial_\mu\phi\partial^\mu\phi\cr
&+ {m^2\over N_C} N_m \int d^2x
Tr(e^{-i\sqrt{4\pi\over N_C N_F}\phi}\tilde gl+
e^{+i\sqrt{4\pi\over N_C N_F}\phi}
\tilde l^\dagger g^\dagger )\cr} \eqno{\eq} $$
with $ m $  given by:
$$ m =[N_Ccm_q({e_c\sqrt{N_F}\over \sqrt{2\pi}})^{\Delta_{C}}
]^{1\over 1+\Delta_{C}} \eqno{\eq}$$
here $\Delta_C$, the dimension of $\tilde h$, is ${N_C^2-1\over
N_C(N_C+N_F)}$.
For the $l=1$ sector, defining
$g^{'}= \tilde ge^{-i\sqrt{4\pi\over
N_C N_F}\phi}\subset U(N_F)$ one gets the
effective action
$$ S_{eff}[g^{'}]
=N_CS[g^{'}]+m^2N_m\int d^2x Tr_F(g^{'}+g^{'\dagger})\eqno{\eq}$$
Thus, the low energy effective action in the $l=1$ sector coincide with the
result of ``naive" approach of the product scheme given by eqn. (4.15).

 \endpage
\def\bx{\vcenter{\hrl\kern 0.0pt
\hbox{\vrl\kern 0pt\vbox{\hbox{\phantom{\vrule
height 8pt depth 4pt width 12pt}}
\kern 0.0pt}\kern 0pt\vrl}\hrl}}
\def\hrl{\hrule}\def\vrl{\vrule}

\chapter { The Baryonic Spectrum of Multiflavor $QCD_2$}
\sect {Classical Soliton solutions}
\par We now look for
static solutions of the classical action.
For a static field
configuration
\footnote{\dagger}{From here on we omit the prime      from
$g'$ so we denote $g\in U(N_F)$} $g(x)$
, the WZ term does not contribute. One
way to see this is
by noting that the variation of the WZ term can be written as
$$\delta {WZ} \propto\int d^2x \varepsilon^{ij}Tr(\delta g)g^\dagger
(\partial_ig)(\partial_jg^\dagger) \qquad       \eqno{\eq}$$
and for $g$ that has only spatial dependence $\delta {WZ}$=0.
Without loss of generality we may take, for the lowest
energy, a diagonal $g(x)$
 \REF\Col{S. Coleman, private communication (1986).}\refend,
$$g(x) = \left( e^{-i\sqrt{{4\pi\over N_C}}
\varphi_1},...,
e^{-i\sqrt{{4\pi\over N_C}}\varphi_{N_F}}\right) \eqno{\eq}$$
For this ansatz and with a redefinition
of the constant term, the action density reduces to
$$\tilde S_d[ g] = -\int dx\sum^{N_F}_{i=1} \left[
{1\over 2}
({d\varphi_i \over dx})^2
-2m^2 \left( cos \sqrt{{4\pi\over N_C}} \varphi_i -1
\right) \right] \eqno{\eq}$$
\par
This is a sum of standard Sine Gordon actions
\REF{\Co}{S. Coleman: ``Classical
lumps and their quantum descendants",
Lectures in the International
School of Subnuclear Physics Ettore Mayorana Erice 1975.}\refend\
For each $\varphi_i$ the well
known solutions of the associated equations of motion are:
$$\varphi_i (x)  = \sqrt{{4N_C \over \pi}}arctg
[ e^{(\sqrt{{8\pi\over N_C}}mx)} ] \eqn{\mishbg}$$
with the corresponding classical energy,
$$E_i = 4m\sqrt{{2N_C\over \pi}} \qquad i=1,....N_F
\eqn{\mishbee}$$
Clearly the minimum energy configuration for this class of ansatz
is when only
one $\varphi_i$ is nonzero, for example:
$$g_\circ(x) = Diag(1,1,....,e^{-i\sqrt{{4\pi\over N_C}}
\varphi(x)}) \qquad \cdot \eqno{\eq}$$
\par
Conserved charges, corresponding to the vector current eqn.(3.11),
 can be computed using the
definition:
$$Q^A[g(x)] = {1\over 2}\int \ dx
Tr(J_0T^A), \eqn{\mishbe}$$
where T$^A$/2 are the $SU(N_F)$ generators  and the U(1)
baryon number is generated by the unit matrix (instead
of ${1\over 2} T^A$ in the equation).  This follows
from $J_\mu = J^A _\mu T^A$, and in the fermionic
basis $J^A _\mu = \bar \psi \gamma_\mu {1\over 2} T^A \psi$.
\par
In particular, for eqn.\mishbg\ ,  we get charges
different from zero only for $Q_B$
and $Q_Y$ corresponding
to baryon number and ``hypercharge" respectively:
$$Q^\circ _B = N_C \qquad Q^\circ _Y = -{1\over 2}
\sqrt{{2(N_F-1)\over { N_F }}}N_C, \eqno{\eq}$$
these charges are determined solely by
the boundary values of $\varphi (x)$, which are:
$$\sqrt{{4\pi\over N_C}}\varphi (\infty) = 2\pi \qquad
\sqrt{{4\pi\over N_C}} \varphi (-\infty) = 0 \qquad \cdot\eqno{\eq}$$
\par
Under a general U$_V(N_F)$
global transformation
$g_\circ(x) \rightarrow \tilde g_\circ (x)
=Ag_\circ(x)A^{-1}$ the energy
of the soliton is obviously unchanged, but
charges other than $Q_B$ and
$Q_Y$ will be turned on.
Let us introduce a parametrization of A that will
be useful later,
$$A = \pmatrix{
& & & & &z_1 \cr
& & &A_{ij} & &\vdots \cr
& & & & &\vdots \cr
& & & & &z_{(N_F-1)} \cr
&Y_1 &\ldots &\ldots &Y_{(N_F-1)} &z_{N_F} \cr} \qquad \cdot
\eqn{\mishbr}$$
Now
$$\tilde g_\circ = 1+(e^{-i\sqrt{4\pi \over N_C}
\varphi}-1)\hbox{z} \eqno{\eq}$$
where (z)$_{\alpha \beta}=z_\alpha z_\beta ^*$ and from unitarity
$\sum^{N_F} _{\alpha=i} z_\alpha z_\alpha ^*$=1.
The charges with $\tilde g_\circ (x)$ are:
$$(\tilde Q^\circ)^A ={1\over 2}N_C
Tr(T^A \hbox{z}) \qquad \cdot\eqno{\eq}$$
Only the baryon number is unchanged. The discussion of the possible
 $U(N_F)$ representations is clearly irrelevant here, since we are
dealing so far with a classical system.
We will return to the question
of possible representations
after quantizing the system.
\sect {Semi-classical quantization and the Baryons}
\par
The next step in the semiclassical analysis is to consider
configurations of the
form
$$g(x,t)=A(t) g_\circ(x) A^{-1}(t) \qquad A(t) \in U(N_F),
\eqn{\mishbs}$$
and to derive the effective action for A(t). Quantization of this
action corresponds to doing the functional integral over g(x,t) of the
above form. The effective action for A(t) is derived by substituting
$g(x,t)=A(t)g_\circ(x)A^{-1}(t)$ in the original action.
\REF{\VDP}{P. di Vecchia, B. Durhuus,
and J.L. Peterson, Phys. Lett. {\bf 114B}
(1984) 245;
D. Ganzales and A.N. Redlich, Phys. Lett.
{\bf 147B} (1984) 150.}
Here we use the following property of the WZ action\refmark{\Fh,\VDP}
$$S\left[ AgB^{-1} \right]=S\left[ AB^{-1}
\right] + S\left[ g,\tilde A_\mu \right] \eqno{\eq}$$
where $S$[g] and $S$[g,$\tilde A$]
are given by equations (3.2) and \mishbd\
respectively,  with the gauge field $\tilde A _\mu$
given as:
$$i\tilde A _+ = A^{-1} \partial_+ A \qquad
i\tilde A_- = B^{-1}\partial_-B \qquad A,B \in U(N_F)
\qquad \cdot \eqno{\eq}$$
Using the above formula for A=B, noting that  S(AA$^{-1}$=1)=0, and
taking $A=A(t)$
$$\partial_+ A =
\partial_- A = {\dot A \over \sqrt 2},
\eqno{\eq}$$
we get
$$\eqalign{&\tilde S\left[ A(t) g_\circ(x)
A^{-1}(t) \right] - \tilde S[ g_\circ ] =\cr
& {N_C \over 8\pi} \int d^2 x Tr
\left\{ [ A^{-1} \dot A, g_\circ ] [ A^{-1}
\dot A,g_\circ ^\dagger ] \right\} \cr
&+ {N_C \over 2\pi} \int d^2xTr \left\{ (A^{-1} \dot A)
(g_\circ ^\dagger
\partial_1 g_\circ) \right\} \cr} \eqn{\mishbh}$$
\par
This action is invariant under global $U(N_F)$ transformations
$A \rightarrow UA$
where
\bk
U $\in$  $G=U(N_F)$. This
corresponds to the invariance of the original
action under
\bk
g $\rightarrow$ UgU $^{-1}$. On
top of that it is also invariant under the local changes
A(t) $\rightarrow$
A(t)V(t) where V(t)$\in$ H= SU($N_F$-1)x U$_B$(1)xU$_Y$(1)
with the last two U(1)
factors corresponding to baryon number and hypercharge, respectively.
This subgroup H of G is nothing but the invariance group of
g$_\circ$(x). In
terms of g$_\circ$(x)
and A(t) the charges associated with the global $U(N_F)$
symmetry, eq. \mishbe, have the following form:
$$Q^B = i{N_C \over 8\pi} \int
dx Tr \left\{ T^B A \left( \left( g_\circ ^\dagger
\partial_1 g_\circ
-g_\circ \partial_1 g^\dagger _\circ \right)
+\left[ g_\circ, \left[ A^{-1} \dot A, g_\circ ^\dagger
\right] \right] \right) A^{-1} \right\} \eqn{\mishbi}$$
\par
The effective action eq. \mishbh\
is an action for the coordinates describing the
coset-space
$$\eqalign{G/H &=
SU(N_F)\times U_B(1)/SU(N_F-1) \times U_Y(1) \times U_B(1) \cr
&= SU(N_F)/SU(N_F-1) \times U_Y(1) = CP^N\qquad \cdot\cr}$$
To see this explicitly we define the Lie algebra valued
variables $q^A$
through $A^{-1} \dot A = i\sum T^A \dot q^A$.  In
terms of these variables \mishbh\ takes the form
(the part that depends on $q^A$):
$$\eqalign{S_q &= \int dt \left[ {1\over 2M} \sum^{2(N_F-1)} _{A=1}
(\dot q^A)^2 -N_C
\sqrt{{2(N_F-1) \over N_F}} \dot q^Y \right] \cr
{1 \over 2M} &= {N_C \over 2\pi} \int^\infty _{-\infty}
(1-cos \sqrt{4\pi \over N_C}\varphi)dx={\sqrt 2 \over m}
({N_C \over \pi})^{3/2}
\cr} \eqn\mishesq$$
The sum is over those $q^A$ which correspond
to G/H generators and $q^Y$
is associated with
the hypercharge generator.
Although the $q^A$ seem to be a ``natural"
choice of variables for the action eq. \mishbh, which
depends only on the combination
$A^{-1} \dot A$, they are not a
convenient choice of variables. The reason for that is the explicit
dependence of the charges \mishbi\
on $A^{-1}$(t) and A(t) as well as on $A^{-1} \dot A$(t).
\par
Instead we found that a
convenient parametrization is that of \mishbr \ .
One
can rewrite the action \mishbh, as well as
the charges \mishbi, in terms of the $z_1$,........,$z_{N_F}$
variables, which however are
subject to the constraint
$\sum ^{N_F} _{\alpha=1} z_\alpha z^*_\alpha$ =1. Thus
$$\tilde S\left[ A(t) g_\circ A^{-1} (t) \right] -
\tilde S[ g_\circ
] =S[ z_\alpha (t), \varphi(x) ] \eqno{\eq}$$
where
$$\eqalign{S\left[ z_\alpha (t), \varphi(x)\right] &=
{N_C \over 2\pi} \int d^2x \{ (1-cos \sqrt{{4\pi
\over N_C}} \varphi)
[ \dot z^* _\alpha \dot z_\alpha \cr
&- (z^* _\gamma \dot z _\gamma)
(\dot z ^* _\beta z_\beta) ]
-i \sqrt{{4\pi \over N_C}} \varphi' z_\alpha^* \dot z _\alpha \}
\cr} \eqn\mishes$$
We can do the integral over
$x$ and rewrite \mishes\ as follows:
$$S[ z_\alpha(t)] = {1\over 2M}
\int dt[ \dot z^* _\alpha \dot z _\alpha - (z^* _\gamma
\dot z_\gamma)(\dot z^* _\beta z_\beta)]
-i{N_C \over 2} \int dt (z^* _\alpha \dot z_\alpha
-\dot z^* _\alpha z_\alpha) \eqn{\mishbrr}$$
where 1/M is defined in equation \mishesq.
The first term in \mishbrr\  is the
usual CP$^{N_F-1}$ quantum mechanical
action, while the second term is a modification due to the {WZ} term,
as obtained from eq.(4.16).
Similarly we express
the $U(N_F)$
charges in terms of the z variables, using equation \mishbi\ :
$$\eqalign{Q^C &= {1\over 2} T^C _{\beta \alpha} Q_{\alpha\beta} \cr
Q_{\alpha \beta}&=N_C z_\alpha z^*_\beta + {i \over 2M}
[ z_\alpha z^*_\beta (z^* _\gamma \dot z_\gamma
-\dot z^* _\gamma z_\gamma) + z_\alpha \dot z^* _\beta
-z^*_\beta \dot z_\alpha ] \cr} \eqn{\mishbj}$$
Of course the symmetries of S[z] are
the global $U(N_F)$ group under which
$$z_\alpha\rightarrow z'_{\alpha}=U_{\alpha\beta}z_\beta
\qquad U\in U(N_F)\eqn{\mishbk}$$
and a local U(1) subgroup of H under which:
$$z_\alpha\rightarrow z'_{\alpha}=e^{i\delta(t)} z_\alpha
\qquad\cdot\eqn{\mishbm}$$
As a consequence of the gauge invariance one can rewrite the action in a
covariant form
 $$S[z_{\alpha}]\,=\,{1\over 2M}\,\,\int dt \Tr{}
(Dz)\sp{\dagger}Dz\,+\,{i}N_C\,\int dt\Tr{}\dot z \sp{\dagger}z
,\eqn\mishbrra$$ where
$$(Dz)_{\alpha}\,=\,\dot z_{\alpha}\,+\, z_\alpha
(\dot z\sp{\star}_{\beta} z_\beta
),\eqn\mishDz$$

Constructing Noether charges
of the $U(N_F)$ global invariance of \mishbk\  out of the action (5.26)
leads to expressions identical with \mishbj\ .
Note that in eqn. \mishDz\ we can view
$\dot z\sp{\star}_{\beta} z_\beta= ia(t)$
as a composite $U(1)$ gauge potential.
\par
Now let us count the
degrees of freedom. The local U(1) symmetry allows us to take one of
the z's to be real,  and the constraint
$\sum _\alpha z_\alpha z_\alpha ^*$=1
 removes one more degree
of freedom, so altogether we are left over with
$ 2N_F-2=2(N_F-1)$  physical
degrees of freedom. This is exactly the
dimension of the coset-space $SU(N_F)\over SU(N_F-1)\times U(1)$.  The
corresponding phase space should have real dimension of $4(N_F-1)$.
Naively, however, we have a phase space of $4N_F$ dimensions and,
therefore, we expect 4 constraints.
\REF{\RSY}{E. Rabinovici, A. Schwimmer and S. Yankielowicz,
\np{248} (1984) 523.}
There are several methods of quantizing systems with constraints.
Here we choose to
eliminate the redundancy in the $z$  variables  and then invoke the
canonical
quantization procedure. (For a different procedure see
[\RSY ]).
But before following these lines let us
briefly describe another method, through
the use of  Dirac's brackets.
We outline the classical case. The quantum case is obtained by
replacing $\{\ ,\ \}$ with $i
[\ ,\ ]$.
The first step in this prescription is to  add to the Lagrangian
  a term of the form
  $\lambda(\sum _\alpha z_\alpha z_\alpha ^*-1)$, in which case
   the conjugate momentum
$\Pi_\lambda$
   of the Lagrange multiplier
vanishes. By
requiring that this condition be preserved in time
one get the constraint $\Phi_1 =(\sum _\alpha z_\alpha z_\alpha ^* -1)=0$.
further  imposing $\dot \Phi_1 = \{\Phi_1, H\}_P= 0$, where $\{\ \}_P$
denotes a Poisson bracket,
one finds     another second
class constraint $\Phi_2 = \Pi\cdot  z + z^\dagger \cdot \Pi^\dagger$. In
addition there is a first class constraint $\Phi_3 = \Pi\cdot  z - z^\dagger
\cdot \Pi^\dagger$,
which corresponds to the local $U(1)$ invariance of the model.
Fixing this symmetry one  gets an additional constraint $\Phi_4$.
For instance one can
 choose the unitary gauge $\Phi_4 =
z_{N_F}-z^*_{N_F}$.
The next step is to compute the constraint matrix
$\{ \Phi_i,\Phi_j\}_P = c_{ij}$.
In the constrained theory, the brackets between
$F$ and $G$ are replaced by  the Dirac brackets of those operators,
given by :
$$\{ F, G\}_D = \{F, G\}_P -\{F,\chi_i\}_P(c_{ij}^{-1})
\{\chi_j, G\}_P\eqn\misheek$$
where  $c^{-1}_{ij}$ is the inverse of the constraint matrix.
 Imposing the constraints  as operator relations it is easy to see that
 $z_{N_F} , \Pi_{N_F}$ and their complex conjugates can be eliminated.
The brackets    for the rest of the fields coincide
\REF{\DMV}{A.C. Davis, A.J. Macfalane, and J.W. Van Holten, Nucl. Phys.
{\bf B232} (1984) 473.}\refend\
with  the results we derive below, when eliminating the constraints
explicitly.

We now describe in some details the quantization of the system
 using unconstrained variables.
We want to choose a set of new variables so
that the constraint
$\sum_{\alpha=1} ^{N_F} z_\alpha z_\alpha ^*$=1 is
automatically fulfilled. There
is a standard choice of such variables,
namely\REF{\CS}{E. Cremer and J. Sherk, Phys. Lett. {\bf 74B} (1978) 341.}
\refend\
$(for\ i=1,....N_F-1)$
$$\eqalign{z_i &={k_i\over
\sqrt{1+X}}\qquad z^*_i ={ k^*_i\over
 \sqrt{1+X}}\ \ \ \
z_{N_F}={e^{i\chi}\over
\sqrt{1+X}}\cr
\hbox{where} \ \ \  \ \ \ \ \ \
X &= \sum^{N_F-1}_{i=1} k^*_i k_i \cdot\cr} \eqno{\eq}$$
The $k_i, k^*_i$ and $\chi$ are $2N_F-1$
real variables with no constraints on them.
The phase space will now have
dimension $2(2N_F-1)$ and we still have extra two
constraints. After
some straightforward algebra we can
write\refmark\DMV
$$\eqalign{S[ k,k^*,\chi ] &= \int dt L(k,k^*, \chi) \cr
L(k,k^*,\chi) &= {1\over 2M} \dot k^*_i h_{ij} \dot k_j -
i{N_C \over 2} \ {{k_i^* \dot k_i - \dot k_i^* k_i} \over {1+X}} \cr
&+ {1\over 2M} {{X \over {(1+X)^2}}} \dot \chi^2
+ \dot \chi \{ {i \over 2M} {{k^*_i \dot k_i - \dot k_i^* k_i}
\over {(1+X)^2}} + {N_C \over {1+X}} \} \cr} \eqno{\eq}$$
where
$$h_{ij} = {\delta_{ij}\over {1+X}} - {{k_i k^*_j} \over
{(1+X)^2}} \qquad \cdot\eqno{\eq}$$
The local U(1) transformations of the z variables transcribe into the
transformations
$$\delta \chi = \epsilon(t); \ \ \delta k_i =
i\epsilon(t) k_i; \ \
\delta k^*_i = -i\epsilon(t) k^*_i \eqno{\eq}$$
and $\delta L= -N_C \dot \epsilon$
just  as in terms of the z variables. This local U(1)
symmetry can be made manifest by defining the covariant derivatives
$$Dk_i = \dot k_i - i\dot \chi k_i \qquad
Dk^*_i = \dot k^* _i + i\dot \chi k^*_i \eqno{\eq}$$
The lagrangian can then be recast in a manifestly gauge invariant form:
$$L(k,k^*,x)={1 \over 2M} Dk^*_i
h_{ij} Dk_j -i{N_C \over 2} \
{{k^*_i Dk_i -(Dk^*_i)k_i}
\over {1+X}} + N_C \dot \chi \qquad \cdot \eqn{\mishbn}$$
Although one can now fix the gauge, for instance
$\dot \chi=0$,  we will continue to work with \mishbn\ .  The
conjugate momenta are given by
$$\eqalign{
\pi_i =    {\partial L \over \partial \dot k_i}
=&{1 \over 2M} Dk^* _j h_{ji} - i{N_C \over 2} {k^*_i \over
1+X} \cr
\pi^*_i= {\partial L \over \partial\dot k_i^*}
=&{1\over 2M} h_{ij} \ Dk_j + i{N_C \over 2} {k_i \over 1+X}
 \cr
\pi_\chi =    {\partial L \over \partial \dot \chi}
=&{i\over 2M} (k^*_i h_{ij} Dk_j - Dk^*_i h_{ij} k_j) +
N_C {1\over 1+X}\cr} \eqno{\eq}$$
Since $h_{ij}$ is invertible we can solve for
$Dk^*_i,Dk_i$ in term of the phase space
variables
$$\eqalign{Dk^*_i &= 2M [ \pi_j + i{N_C \over 2}
{k^* _j\over 1+X} ] h^{-1} _{ji} \cr
Dk_i &= 2M h^{-1} _{ij} [ \pi^*_j - i{N_C \over 2}
{k_j \over 1+X} ] \cr} \eqno{\eq}$$
where
$$h_{ij}^{-1} = (1+X)(\delta_{ij} + k_i k^*_j) \eqno{\eq}$$
Also
$$\pi_\chi = i(k^*_i \pi^*_i - \pi_i k_i) + N_C \eqno{\eq}$$
giving the constraint equation
$$\psi = \pi_\chi -i(k^*_i \pi^*_i - \pi_i k_i) -N_C = 0 \eqno{\eq}$$
The canonical Hamiltonian is given by
$$\eqalign{H_c &= \pi_i \dot k_i + \pi^*_i \dot k^*_i
+\pi_{\chi}\dot\chi -L  \cr
&= 2M [ \pi_i + i{N_C k^*_i \over 2(1+X)} ]
h_{ij} ^{-1} [ \pi^*_j - i{N_Ck_j \over 2(1+X)} ]\cr
&+ \dot \chi [ \pi_\chi -i(\pi^*_i k^*_i - \pi_i k_i) - N_C ] \cr}
\eqno{\eq}$$
and this can be further simplified to:
$$\eqalign{H_c &= 2M(1+X) [ \pi_i \pi^*_i + (\pi_i k_i)
(\pi^*_i k^*_i) \cr
&- i{N_C \over 2}(\pi_i k_i - \pi^*_i k^*_i) +
{1\over 4}{N_C ^2 X \over (1+X)} ] + \dot \chi\psi \cr}
\qquad \cdot \eqno{\eq}$$
Here H$_c$ is obtained explicitly in terms of the canonical variables
$k_i,k^*_i,\pi_i,\pi^*_i$.
The $\dot \chi\psi$ term indicates that
$\dot \chi$ also behaves as a Lagrange
multiplier since,
 following the Dirac procedure, we should define
$$H_T = H_c + \lambda(t) \psi \eqno{\eq}$$
where $\lambda$ is a priori an
arbitrary function of $t$.  We could absorb the $\dot \chi$ in
$\lambda$.
\par
Quantization of this Hamiltonian is now essentially straightforward.
Let us first consider the symmetry generators
$Q_{\alpha\beta}$, which in terms of the new canonical variables
take the form
$$\eqalign{Q_{ij} &= i(k_i \pi_j - \pi^*_i k^*_j) \cr
Q_{i,N_F} &= e^{-i\chi} [ {N_C k_i \over 2} - i
(\pi^*_i + k_i \pi_j k_j) ] \cr
Q_{N_F,i} &= e^{i\chi} [ {N_C k^*_i \over 2} + i
(\pi_i +  k^*_j\pi^*_j  k^*_i
) ] = Q^* _{i,N_F}\cr
Q_{N_F,N_F} &= N_C -i(\pi_i k_i - \pi^*_i k^*_i) \cr}
\eqn{\mishbo}$$
We will now show that the
H$_T$ can be expressed in terms of the second
Casimir operator of the $SU(N_F)$ group.
\par
The second $U(N_F)$ Casimir operator is
related to charge matrix elements
$Q_{\alpha \beta}$ in the following way:
$$Q_A Q^A = {1\over 2} Q_{\alpha \beta} Q_{\beta
\alpha} \eqno{\eq}$$
A straight forward substitution gives:
$$\eqalign{{1\over 2} Q_{\alpha \beta}
Q_{\beta \alpha} &= (1+X) [ \pi^*_i \pi_i +
\pi_i k_i \pi^*_j k^*_j \cr
&-i{N_C \over 2} (\pi_i k_i - \pi^*_i k^*_i)] + {1\over 2}
N_C^2 (1+{X\over 2})\cr} \eqno{\eq}$$
Therefore, the Hamiltonian is:
$$H_T = 2M [ Q^A Q^A - {N_C^2 \over 2} ]
+ \lambda (t) \psi \eqn{\mishbrrr}$$
Denoting the $SU(N_F)$ second Casimir operator
by $C_2$, and using
$Q_A Q^A = C_2 + {1\over 2N_F}
(Q_B)^2$ we get:
$$H_T = 2M [ C_2 - N_C^2 {(N_F-1)\over {2N_F}} ] \eqn{\mishbq}$$
\par
The fact that H$_T$ is,  up to a constant,
the second Casimir operator, is
another way to show that the
charges $Q_{\alpha \beta}$
are conserved. These conserved charges will generate
symmetry transformations via:
$$\eqalign{\delta k_i &=i[ Tr(\epsilon Q),k_i]
\qquad \delta k^*_i =i[ Tr(\epsilon Q),k^*_i)] \cr
\delta \chi &=i[ Tr(\epsilon
Q),\chi] \cr} \eqn{\mishbp}$$
and similar equations for the momenta
$\pi_i, \pi^*_i, \pi_\chi$. Here
$\epsilon_{ij} = {1\over 2}\epsilon^A T^A _{ij}$ is the
matrix of parameters. The transformation
laws are derived using the
constraint equation $\psi$=0
after performing the commutator calculations.
Notice that $Q_{ij}$
and $Q_{N_F,N_F}$ are linear
in coordinates and momenta and
therefore, the $SU(N_F-1)\times U_Y(1)$
transformations they generate are linear.  The
$Q_{N_F,i}$
and $Q_{i,N_F}$ charges, on the other hand, have cubic terms as well
(quadratic in coordinates),
so that the coset-space transformations
of $SU(N_F)\over SU(N_F-1)\times U(1)$ are non-linear.
This is a well known property of CP$^n$ models.
Substitution of $Q_{\alpha \beta}$ in equation \mishbp\  gives:
$$\delta k_l=i[ \epsilon_{ji} k_i \delta_{jl}
+e^{i\chi} \epsilon_{iN_F} \delta_{il}-e^{-i\chi}
\epsilon_{N_Fi} k_i k_l - \epsilon_{N_FN_F} k_l]
\eqno{\eq}$$
where we used
$i[ k,\pi]$=1. Inversely, starting with
 these transformation laws it is easy to verify the invariance of
the action. The standard Noether procedure then
gives the charges $Q_{\alpha \beta}$
in terms of the coordinates and velocities,
which (not suprisingly) coincide with those
given in equation \mishbo\ .
One could also deduce these transformation laws by making the
change of variables
$z_\alpha, z^*_\alpha \rightarrow k_i, k_i^*, \chi$ in \mishbj\  directly.
\par
One  can  verify that
$$[ Q^A, Q^B ] = if^{ABC} Q^C \eqno{\eq}$$
where $f^{ABC}$ are the structure constants of the
$U(N_F)$ group.
\par
Do we have further restrictions on the physical states? We
shall see now
that in fact we do have. Remember that our lagrangian \mishbn\
includes an
auxiliary gauge field $A_\circ \equiv\dot \chi$ and thus
 has to obey the associated Gauss law:
$${\partial L \over \partial A_\circ} = {\partial L \over
\partial \dot \chi} =
\pi_\chi = N_C - i (\pi_i k_i - \pi_i ^* k_i ^*)
=0 \eqno{\eq}$$
Since $\pi_\chi$ is a linear combination of
$Q_B$ and $Q_Y$, and the first is
constrained to be $Q_B=N_C$, the $Q_Y$
is restricted as well. More
specifically, $Q_Y=\bar Q_Y$, with
$$\bar Q_Y = {1\over 2} \sqrt{{2 \over (N_F-1)N_F}} N_C
\qquad \cdot \eqno{\eq}$$
\sect{The baryonic spectrum}
\par
The masses of the baryons \mishbee\  and \mishbq,   and the two constraints on
the multiplets of the physical states, namely
$Q_B=N_C$ and that the multiplets contain
$Q_Y=\bar Q_Y={1\over 2} \sqrt{{2\over (N_F-1)N_F}}N_C$,
are the main results of the last section.
All states of the multiplet with
$Q_Y \ne\bar Q_Y$ will
be generated from the state $Q_Y=\bar Q_Y$
by $SU(N_F)$ transformations as in eq. \mishbs \ .
Using the above constraints we can investigate now what
possible representations will
appear in the low energy baryon sector.
Considering states with quarks only (no antiquarks), the
requirement of $Q_B=N_C$ implies that only
representations described by Young
tableau with $N_C$ boxes appear. The extra constraint
$Q_Y=\bar Q_Y$ implies that all $N_C$ quarks
are from $SU(N_F-1)$, not involving the $N_F$th.
These are automatically obeyed in the totally symmetric representation
of $N_C$ boxes.
\REF\us{Y. Frishman and W.J. Zakrzewski, \np
{328} (1989) 375.}
\REF\uss{Y. Frishman and W.J. Zakrzewski, \np
{331} (1990) 781.}
In fact, this is the only representation possible for
flavor space,\refmark\us
since the states have to be constructed out of the
components of one complex vector $z$ as $\prod_{i=1}^{N_F}z_i^{n_i}$ with
$\sum_i n_i = N_C$.
See also more detailed discussion in the next section. For another way
of deriving this result  see section (5.5).
Thus for $N_C=3, N_F=3$ we get only $10$ of
$SU(3)$. This is understandable, since there is no physical spin in two
dimensions.
\par
What about the masses of the baryons? The total mass of a baryons
is given by the sum of \mishbee\  and \mishbq \  namely
$$E=4m \sqrt{{2N_C \over \pi}} + {m \sqrt 2}
\sqrt{{\left({\pi\over N_C}\right)}^3} \left[
C_2 - N_C^2 {(N_F-1)\over 2N_F} \right] \qquad \cdot \eqn{\mishE}$$
\par
For large $N_C$, the classical term behaves like $N_C$, while the
quantum correction like $1$. This will be worked out in section (5.5).
That the total mass goes like $N_C$ for  large $N_C$, and that the
quantum fluctuations are ${1\over N_C}$ of the classical result, is in
accord with general considerations (see ref.
\REF\Wln{E. Witten, \np {160}
(1979) 57.}
[\Wln ] ).


\REF{\DonoNappi}{ J. Donoghue and C. Nappi,
\pl{168} (1986) 105.}

\REF{\YabuKK}{H. Yabu \journal Phys. Lett. &B218 (89) 124;
D. B. Kaplan and I. Klebanov, \np {335} (1990) 45.}

\REF{\Weigel}{H. Weigel, J. Schechter, N.W. Park and Ulf-G. Meissner,
\prd {42} (1990)  3177.}

\sect{Flavor quark content of the baryons}
A measure of the
 quark content of a given flavor $q_i$ in a baryon state
$\ket{B}$ is given by
$$\VEV{\bar{q}_i q_i}_B =
\int d x \VEV{g_{ii}}_B-
\int d x \VEV{g_{ii}}_0$$
$$=\int d x z_i^* z_i
\VEV{\lsquare e^{-i\sqrt{{4\pi\over N_C}}\phi_c}-1\rsquare}_B
\eqn\qbarqDef$$
$$=\hbox{const.} \VEV{z_i^* z_i}_B$$

In order to make contact with the real world, we take here
$N_C=3$ and $N_F
=3$, getting the baryons in the
 $\bold{10}$ representation of flavor.
Similarly, for $SU_F(2)$ there is only the isospin ${3\over 2}$
representation.\refmark{\Gepner}
This is what we would expect from \naive\ quark model considerations.
The total wave function must be antisymmetric. Baryon is a color
singlet, so the wavefunction is antisymmetric in color and it must
be symmetric in all other degrees of freedom. There is no spin, so
the baryon must be in a totally symmetric representation of the
flavor group, a $\bold{10}$ for three flavors.\refmark\rMarek
Therefore,  strictly speaking there is no state analogous to the
proton. On the other hand, there is a state which is
the analogue of the $\Delta^+$, namely the charge 1 state in the
$\bold{10}$ representation,  $z_1^2 z_2$.
The $\bold{10}$
 is the lowest baryon multiplet in QCD$_2$.
In the following we shall be dealing with the relative weight
of a given flavor in some baryon state .
Thus, $\VEV{\bar{q} q}_B$ will henceforth stand for the ratio
$$ { \VEV{\bar {q} q}_B \over
     \VEV{\bar {u} u + \bar{d} d + \bar{s} s }_B}$$

For $\Delta^+\sim z_1^2 z_2$ we obtain\refmark\rMarek

$$\VEV{\bar{s} s}_{\Delta^+}=
{\displaystyle
\int d^2 z_1 d^2 z_2 \vert z_3\vert^2 (z_1^2 z_2) (z_1^2 z_2)^*\over
\displaystyle
 \int d^2 z_1 d^2 z_2         (z_1^2 z_2) (z_1^2 z_2)^* }
={1\over 6}\eqn\ssDplus$$
as well as
$$\VEV{\bar{u} u}_{\Delta^+}={1\over 2}\qquad\ \ \VEV{\bar{d}
d}_{\Delta^+}={1\over 3}\eqn\ddDplus$$ %
In evaluating the integral in the numerator in eq.~\ssDplus\
we have used
$\vert z_3\vert^2=1-\vert z_1 \vert^2 - \vert z_2 \vert^2$, which
follows from the unitarity of the matrix $A$ in eq. \mishbs\
Similarly, for $\Delta^{++}\sim z_1^3$ we have
$$\VEV{\bar{u} u}_{\Delta^{++}}={2\over 3}\qquad\ \
\VEV{\bar{d} d}_{\Delta^{++}}={1\over 6}\qquad \ \ %
\VEV{\bar{s} s}_{\Delta^{++}}={1\over 6}\eqn\udsDpp$$
In the constituent quark picture $\Delta^{++}$ contains
just three $u$ quarks. Both the $d$-quark and the $s$-quark
content of the $\Delta^{++}$ come only from virtual quark
pairs. Therefore in the $SU(3)$-symmetric case
$\VEV{\bar{s} s}_{\Delta^{++}}=
 \VEV{\bar{d} d}_{\Delta^{++}}$, and
$\VEV{\bar{s} s}_{\Delta^{+}}=
 \VEV{\bar{s} s}_{\Delta^{++}}$, as expected.

{}From eqn. \hbox{\udsDpp }
one can also read the results for
$\Omega^-\sim z_3^3$, by replacing $u\leftrightarrow s$.
In the general case of $N_F$ flavors and $N_C$ colors,
one obtains\refmark\rMarek
$$\VEV{(\bar{q}q)_{sea}}_B={1\over N_C+N_F},\eqn\GenCaseSea$$
where
$(\bar{q}q)_{sea}$ refers to the non-valence quarks in the
baryon B. Moreover,
one can also compute flavor content of valence
quarks. Consider a baryon $B$ containing $k$ quarks
of flavor $v$. The $v$-flavor content of such a baryon is
$$\VEV{\bar{v} v}_B={k+1\over N_C + N_F} \eqn\GenCaseV$$
This implies an ``equipartition" for valence and sea,
each with a content of $1/(N_C+N_F)$. It also follows that
the total sea content of $N_F$ flavors is
$$\sum_{q=1}^{N_F}\VEV{(\bar{q}q)_{sea}}_B={N_F\over N_C
+N_F}
\eqn\TotSea$$
which goes to zero for fixed $N$ and $N_C\rightarrow\infty$,
as expected.

It is interesting to compare these results with the Skyrme
model in 3+1 \break
dimensions.\refmark{\DonoNappi-\Weigel}
For the proton\refmark{\DonoNappi}
$$\VEV{\bar{u} u}^{3+1}_{p}={ 2\over 5}\qquad\ \ \VEV{\bar{d}
d}^{3+1}_{p}={11\over30}\qquad\ \  \VEV{\bar{s} s}^{3+1}_{p}={
7\over30}\eqn\udsP$$ %
and for the $\Delta$
\refmark{\Weigel}
$$\VEV{\bar{s} s}_{\Delta    }^{3+1}={ 7\over24}\qquad\qquad
  \VEV{\bar{s}
s}_{\Omega^{-}}^{3+1}={ 5\over12}\eqn\udsD$$

The qualitative picture is similar, although
the $\bar{s} s$ content in the non-strange baryons is lower in
$1+1$ dimensions. One may speculate that in
$1+1$ dimensions the effects of loops are smaller than in
$3+1$ dimensions, since the theory is super-renormalizable
and there are only longitudinal gluons.
In the $SU_F(3)$-symmetric
limit the strange quark content of baryons with zero net
strangeness is significant, albeit smaller than that of
either of the other two flavors. The situation obviously
is reversed for $\Omega^-$.

In the real world the current mass of the strange quark is
much larger than the current masses of $u$ and $d$ quarks.
It is natural to expect that this will have the effect of
decreasing the strange quark content from its value in the
$SU_F(3)$ symmetry limit. We do not know the exact extent
of this effect, but it is likely that the strange content
decreases by a factor which is less than two. This estimate
is based on both explicit model calculations\refmark{\YabuKK-
\Weigel}
and what we know from PCAC, namely that
the analogous quark bilinear expectation
values in the vacuum are not dramatically different from
their $SU(3)$ symmetric values:
$$ 0.5 \lsim  {\VEV{\bar{s} s}_0 \over
\VEV{\bar{u} u}_0} \lsim 1\eqn\vacVEVs$$
\section{Multibaryons}
Let us now explore the possibility of having multi-baryons states.
The procedure follows similar lines to that of the baryonic spectrum,
namely, we look for classical solution of the equation of motions with
baryon number $kN_C$ and then we semiclassically quantize it.
The ansatz for the classical solution
of the low lying
$k$-baryon state is\refmark\YFWZ
  taken now to be
$$g_0(k)\,=\,\pmatrix{\overbrace{\matrix{1&&\cr&\ddots&\cr}
}\sp{(N_F-k)}&\cr &\overbrace{\matrix{
exp[-i({4\pi\over N_C})\sp{1\over 2}\varphi_c]&&\cr&\ddots&\cr}}
\sp{k}&\cr},\eqn \kstate$$
For the semi-classical quantization we generalize the parametrization
given in \mishbr\  to the following
$$A\;=\;\pmatrix{&  &\cr
A_{ij}& &z_{i\alpha}\cr
      &&\cr},\eqn\A$$
where $i$ represents the rows ($1,\ldots,N_F$) and $\alpha$ the columns
($N_F-k+1,\ldots,N_F$).
 The effective action in its covariant form \mishbrra\
   becomes\refmark\YFWZ
 $$S[z_{\alpha}]\,=\,{1\over 2M}\,\,\int dt \Tr{}
(Dz)\sp{\dagger}Dz\,+\,{i}N_C\,\int dt\Tr{}\dot z \sp{\dagger}z
,\eqn\ezact$$
where now, instead of eqn. \mishDz\
$$(Dz)_{i\alpha}\,=\,\dot z_{i\alpha}\,+\,z_{i\beta}(\dot z\sp{\star}_{j\beta}
z_{j\alpha}),\eqn\eDz$$
Using the same steps as those which led to \mishbq\  one finds now
the Hamiltonian\refmark\YFWZ
$$H\;=\;2M\Bigl[C_2(N_F)\,-\,{N_C\sp2\over 2N_F}k(N_F-k)\Bigr]
\,+\,kE_c,\eqn\ef$$
with $E_c$ the classical contribution for one baryon, the first term in
\mishE.

It was  shown in ref. \(\YFWZ\)
that the allowed $k$-baryon states contain $(kN_C)$
boxes in the Young tableaux representation of the flavour group
$SU(N_F)$.  Let us recall that this result followed from the constraint
implied by the local invariance
$$z_{i\alpha}\;\rightarrow\;e\sp{i \delta(t)}\,z_{i\alpha}.\eqn\egauge$$
Performing a variation corresponding to this invariance we find that the
action $S$ changes by
$$\Delta S\;=\,(kN_C)\,\int\,\dot\delta\,dt.\eqn\edelta$$
This means that the $N_z$ number is equal to $(kN_C)$. Thus  for any
 wave function, written as a polynomial in $z$ and $z\sp{\star}$, the
number of $z$'s minus the number of $z\sp{\star}$'s must equal $(kN_C)$.
Note that for $k=1$ the transformation  \egauge\ represents also the
$N_F^{th}$ flavor number. Thus \edelta\
entails that the representation contains a state with $N_C$ boxes of
the $N_F$ flavor,
 and therefore must be the totally symmetric
representation.
\par Now, the effective action \ezact\  is invariant under a larger group
of local transformations. In fact, we have extra
 $(k\sp2-1)$ generators, which
correspond to $SU(k)$ under which \ezact\ is locally invariant.
This can be
 exhibited by defining\refmark\us ``local gauge potentials"
$$\tilde A_{\beta\alpha}(t)\,=\,-(z\sp{\dagger}\dot z)_{\beta\alpha}.
\eqn\epot$$
Then
$$Dz\;=\;\dot z\,+\,z\,\tilde A.\eqn\eg$$
Under the local gauge transformation corresponding to $\Lambda(t)$,
$\tilde A$ transforms as
$$\tilde A(t)\;\rightarrow\;e\sp{i\Lambda}\tilde A e\sp{-i\Lambda}\,+\,
(\partial_te\sp{i\Lambda})\,e\sp{-i\Lambda}.\eqn\lambda$$
Then we have
$$(Dz)_{i\alpha}\;\rightarrow\;(Dz)_{i\beta}(e\sp{-i\Lambda})
_{\beta\alpha}\eqn\edz $$
and so $\Delta S=0$.  If we perform the $U(1)$ transformation \egauge\ we
obtain a contribution \edelta\
 from the Wess-Zumino term, which implies
$N_z=(kN_C)$.
 But due to the larger local symmetry we have more restrictions; they
imply that the allowed states have to be singlets under the above
mentioned $SU(k)$ symmetry.  This is analogous to the confinement
property of QCD, which
tells that, due to the non-abelian gauge invariance, the physical states
have to be colour singlets. Here we have analogous singlet structure
of the $SU(k)$ in the flavour space. Taking a wave function that has
$z$'s only (analogous to quarks only for QCD), it must be of the form
$$\psi_{k}(z)\,=\,\prod_{i=1}\sp{N_C}\Bigl(\epsilon_{\alpha_1...\alpha_k}
z_{i_1\alpha_1}...z_{i_k\alpha_k}\Bigr) ,\eqn\epsi$$
for a  given  set of $1\le i_1,..i_k\le N_F$.
\par The most general state will be then of the form
$$\tilde \psi(z,z\sp{\star})\,=\,\psi_k(z)[
\prod_{\{i,j\}}(z_{i\alpha}\sp{\star}
z_{j\alpha})\sp{n_{ij} }],\eqn\estate$$
and the products are over given sets of indices.\par
Let us now compute the mass of the state represented by \epsi. Using the
explicit
formula from ref.[\uss], we obtain
$$\eqalign{E[\psi_k]\,=&\,2M\bigl[{k(N_F-k)\over
2N_F}N_C\sp2\,+\,{k(N_F-k)\over
2}N_C\cr
-&\,{N_C\sp2\over 2N_F}k(N_F-k)\bigr]\,+\,kE_c\cr
=&\,Mk(N_F-k)N_C\,+\,kE_c.\cr}\eqn\eenergy$$\par
To obtain binding energies, consider our $k$-baryon as built from
constituents $k_r$, such that $k=\sum_rk_r$.  Then
$$\eqalign{B[k\vert k_r] \,&=\,-(MN_C)
\bigl[k\sp2-\sum_ik_i\sp2\bigr]\cr
&=\,-(2MN_C)\sum_{r>s}k_rk_s\cr}\eqn\ebinding$$
When all $k_r=1$, the sum gives us ${1\over 2}k(k-1)$, \ie\ the number of
one-baryon pairs in the $k$-baryon state.  Note that the binding energy
is always negative, thus the $k$-baryon is stable.  The maximal binding
corresponds to the case when all $k_r=1$.

\par Note also that in the $N_C\,\rightarrow\,\infty$ limit, the binding
tends to a finite value, since then
$$\lim_{N_C\rightarrow\infty}\,(2MN_C)\,=\,
(Cme_C)\sp{1\over
2}\bigl({2N_F\over \pi}\bigr)\sp{1\over 4}\pi\sp{3\over 2}.\eqn\elimit$$
\par Let us take as an example an analogue of a deutron \ie\ a di-baryon $k=2$.
Then taking $N_C=3$, $N_F=2$ we find that its representation is a flavour
singlet (this is the limiting case of $k=N_F$). The ratio of the binding
to  twice the baryon mass is given by
$$\epsilon_2\,=\,{1\over 1+{24\over \pi\sp2}}\,=\,0.29.\eqn\etwo$$
For $k=2$, $N_C=3$ and $N_F=3$ we find that the di-baryon is represented
by $\overline{10}$ and the ratio is given by
$$\epsilon_3\,=\,{1\over 2+{24\over \pi\sp2}}\,=\,0.23.\eqn\ethree$$
For general $N_F$ we obtain $$\epsilon_F\,=\,{1\over (N_F-1)+{24\over
 \pi\sp2}}\,=\,{1\over N_F+1.43}.\eqn\egeneral$$
\par
Finally, let us make the following comment. The ratio of the
quantum fluctuations term to the classical term, in the expression
for the mass eq.\eenergy
, is given by
$$ {Quantum \quad Corrections \over Classical \quad Term}
= \left({\pi^2\over 8}\right) {N_F-k \over N_C}.\eqn \ratio $$
Thus, we do not expect our approximations to hold in the
region $ N_F \geq (N_C+1) $. We expect it to start for
$N_C \geq N_F$, and to be good in the region $N_C \gg N_F $.
 \chapter {Summary and Conclusions}
 One of the outstanding problems of high energy physics is the derivation
 of the hadronic spectrum from QCD, the underlying theory.
 A large variety of methods have been  used  to address this question,
 including       lattice gauge  simulations, low energy effective
 Lagrangians like the Skyrme model and chiral Lagrangians, QCD sum rules
 etc.
 In spite of this  major  effort the gap between the phenomenology and
 the basic  theory has been  only partially bridged.
Another direction that has been taken in order
to gain insight into    the problem is
 a lower dimensional analogous system  $QCD_2$.

The study of  two dimensional problems to improve the understanding of
 four dimensional  physical systems
  was found to be fruitful.
   For example  the study of spin systems in two
dimensions shed light on  four dimensional gauge field theories.
Obviously,
physics in two dimensions is simpler than that of the real world since the
underlying manifold is  simpler and since  the number of
degrees of freedom of
each field is smaller. There are some additional
simplifying features in two
dimensional physics. In one space dimension
there is no  rotation symmetry and no angular momentum. The light
cone is disconnected and  is composed of
 left moving and right moving branches.
 Therefore, massless particles are either
 on one branch or the other.
 These two  properties are the  basic building
blocks of the idea of transmutation between
systems of different statistics.
Also, the ultra-violet behaviour is more convergent in two dimensions,
making for instance $QCD_2$ a superconvergent theory.

Bosonization is the formulation of fermionic systems  only in terms of
bosonic variables and fermionization is just the opposite process.
The study of  bosonized physical systems
offers several advantages:\hfill\break
(1) It
is usually easier to deal with commuting fields rather than
anti-commuting ones.\hfill\break
(2) In certain examples like the Thirring model the
fermionic strong coupling regime
turns into the weak coupling one in its bosonic version, the Sine-Gordon
 model.\hfill\break
 (3) The non-abelian
bosonization,  especially in the product scheme,
offers a separation between
colored and flavored degrees of freedom,  which is
very convenient for the analyzing low
lying spectrum.\hfill\break
(4) Baryons composed of $N_C$ quarks are a many-body problem in the
fermion language, while simple solitons in the boson language.
\hfill\break
(5)  One loop
fermionic computations involving the currents turn into tree level
consideration in the bosonized version. The best
known example of the latter are the chiral ( or axial) anomalies.

An important question that  has to  be addressed when applying
bosonization methods is to what extent   are  the
two formulations equivalent.   Let us first
  demonstrate this
  equivalence in simple ``physical" terms for the
 example of  the  zero charge sector
 of a massless Dirac fermion Fock space.
 A wave function of a
 state composed of
 a fermion and its anti-particle having together zero
 fermionic charge and moving in the same direction, never spreads and the
 two particles will never separate.
 They are therefore indistinguishable from a free massless boson.
A more rigorous argument of the equivalence of the fermionic and bosonic
formulations of colored-flavored Dirac fermions was given in terms of
their algebraic structure. It was shown that the
 currents and  energy-momentum tensor of the
two pictures correspond
to the same representation of the Kac-Moody  and  Virasoro
 algebras.
In both cases the energy momentum tensor
is quadratic in the currents. Using current algebra and conformal
symmetry Ward identities guarantee that correlation functions of
currents in the two  formulations  coincide.
The equivalence properties for the massive theory are less obvious.
As reviewed in this article, the bosonization of the mass term in the
``product scheme" failed to reproduce the fermionic correlators of mass
bilinears. The $U(N_C\times N_F)$ scheme, however, is
e free from such a
problem. The equivalence of the gauged theories
was argued  by
integrating the matter degrees of freedom in the bosonic and fermionic
versions.    The derived    effective actions  in the bosonic
description where equal to those of the fermion models in  various
different regularization schemes.

$QCD_2$ was addressed first in the fermionic formulation. In his
seminal work 't Hooft deduced the mesonic spectrum in the large
$N_c$ limit. However, it seems that it is easier to analyze the
baryonic physical states in the bosonic language.
The extraction of the baryonic spectrum and wave functions from the low
energy  effective action is similar to the steps taken in the Skyrme
model. It is worth mentioning again that whereas in the four
dimensional case it is only an approximated model  derived by an
``educational guess",
in two dimension the
action at the strong coupling regime is exact.

In spite of the progress that has been made in the understanding of
$QCD_2$ there are still several interesting open questions. The
incorporation of more complicated mass matrices and  higher order
corrections to the ${m_q\over e_c}\rightarrow 0 $ limit, are
examples of such questions that are intimately related to the
analysis presented in the review.  Possible relations of the
low-energy effective action to some massive two dimensional
integrable models requires further exploration since it  may lead to
the full solution (not semi-classical) of the strong coupled
$QCD_2$. One may attempt to write down the full bosonized standard
model in two dimension namely to incorporate the electroweak
interaction as well. This extension of $QCD_2$ faces the difficulty
of chiral bosonization. Nevertheless, one may gain in this way some
further insight on the real world standard model.
Obviously, the most
interesting task is to extract  useful methods and that are applicable
to  the four dimensional theory. As  mentioned in the
introduction,
bosonization techniques were applied to four dimensional systems,
like monopole induced proton decay
 and fractional charges induced on
monopoles by light fermions. So far not much has been
 achieved in the application to $QCD_4$ .
The identification of hadronic systems that may be approximated by a
bosonized two dimensional systems is still an open question which deserves
further investigation.

\endpage
\refout
\end